\documentclass[aps,prb,superscriptaddress,floatfix,preprint]{revtex4-1}
\usepackage{bm}
\usepackage{graphicx}

\begin{document}

\title{Spin-orbit enabled quantum transport channels 
in a two-hole double quantum dot}

\author{Alex Bogan}
\affiliation{Security and Disruptive Technologies,
National Research Council of Canada, Ottawa, Canada K1A0R6}

\author{Sergei Studenikin}
\email{Sergei.Studenikin@nrc-cnrc.gc.ca}
\affiliation{Security and Disruptive Technologies,
National Research Council of Canada, Ottawa, Canada K1A0R6}
\affiliation{Department of Physics and Astronomy, University of Waterloo, Waterloo, Ontario, Ontario, NL2 3G1, Canada}

\author{Marek Korkusinski}
\email{Marek.Korkusinski@nrc-cnrc.gc.ca}
\affiliation{Security and Disruptive Technologies,
National Research Council of Canada, Ottawa, Canada K1A0R6}

\author{Louis Gaudreau}
\affiliation{Security and Disruptive Technologies,
National Research Council of Canada, Ottawa, 
Canada K1A0R6}

\author{Jason Phoenix}
\affiliation{Security and Disruptive Technologies,
National Research Council of Canada, Ottawa, 
Canada K1A0R6}
\affiliation{Department of Physics and Astronomy, University of Waterloo, Waterloo, Ontario, Ontario, NL2 3G1, Canada}

\author{Piotr Zawadzki}
\affiliation{Security and Disruptive Technologies,
National Research Council of Canada, Ottawa, Canada K1A0R6}

\author{Andy Sachrajda}
\affiliation{Security and Disruptive Technologies,
National Research Council of Canada, Ottawa, Canada K1A0R6}

\author{Lisa Tracy}
\affiliation{Sandia National Laboratories, Albuquerque, NM, 87185, USA}

\author{John Reno}
\affiliation{
Center for Integrated Nanotechnologies, Sandia National Laboratories, Albuquerque, NM, 87185, USA}

\author{Terry Hargett}
\affiliation{
Center for Integrated Nanotechnologies, Sandia National Laboratories, Albuquerque, NM, 87185, USA}

\date{\today}

\begin{abstract}
We analyze experimentally and theoretically the transport spectra
of a gated lateral GaAs double quantum dot containing two holes.
The strong spin-orbit interaction present in the hole subband
lifts the Pauli spin blockade and allows to map out the complete
spectra of the two-hole system.
By performing measurements in both source-drain voltage directions,
at different detunings and magnetic fields, we carry out
quantitative fitting to a Hubbard two-site model accounting
for the tunnel coupling to the leads and the spin-flip relaxation
process.
We extract the singlet-triplet gap and the magnetic field
corresponding to the singlet-triplet transition in the double-hole
ground state.
Additionally, at the singlet-triplet transition we find a resonant
enhancement (in the blockaded direction) and suppression of current 
(in the conduction direction).
The current enhancement stems from the multiple resonance of two-hole
levels, opening several conduction channels at once.
The current suppression arises from the quantum interference of
spin-conserving and spin flipping tunneling processes.
\end{abstract}

\maketitle

\section{Introduction}

%General motivation: hole spins as qubits briefly review

Recently there has been great interest in utilizing hole spins 
as qubits for solid-state quantum information 
applications.~\cite{Scappucci2020}
This qubit choice is motivated by the expected suppression
of hyperfine interactions between the spin of the hole and those
of the crystal lattice nuclei, promising a low-decoherence
platform without isotopic purification.\cite{Burkard2008,Fischer2008}
The second aspect of the hole subband is the strong
spin-orbit interaction (SOI) compared to that of electron solid-state
devices.
The SOI can be harnessed to manipulate the hole spin coherently
in a local fashion, obviating the need for local pulsed magnetic 
fields.~\cite{Bulaev2007,Venitucci2019,Voisin2016,Maurand2016,Watzinger2018,Bogan2019-2} 
Presently, the main technological effort towards implementation
of the hole spin-based qubits is concentrated in several material
platforms:
complementary metal–oxide–semiconductor silicon
devices,\cite{Voisin2016,Maurand2016}
silicon gated devices,~\cite{Li2015}
the Ge/Si hut quantum
wires,~\cite{Watzinger2018,Gao2020,Watzinger2016,Li2018}
Ge/Si core-shell nanowires~\cite{Hu2007,Hu2012}
and nanocrystals,~\cite{Ares2013}
Ge/SiGe planar heterostructures,~\cite{Hendrickx2020,Hendrickx2019}
and GaAs/AlGaAs gated 
devices.~\cite{Wang2016,Wang2016-2,Bogan2019-2,Bogan2019,Tracy2014,Bogan2017,Bogan2018}

Since the Si and Ge crystal lattice possesses inversion symmetry,
these systems exhibit only the Rashba SOI, which can be tuned, 
and in principle even switched off, by appropriate choices of gate
voltages.~\cite{Bulaev2007,Bulaev2005}
In GaAs/AlGaAs planar heterostructures, on the other hand,
both Dresselhaus and Rashba SOIs are present, with the former
determining the spin relaxation time.~\cite{Bogan2019}
In gated dots in which electrons are confined, the effects of the SOI
can be eliminated by appropriate choice of the direction of
the magnetic field, which determines the direction of spin
quantization.~\cite{NadjPerge2010,Nichol2015}
In contrast, the spin of the confined hole is pinned
in the direction of the strongest confinement, i.e.,
the heterostructure growth direction.
Such pinning
results in zero in-plane effective $g^*$-factor,~\cite{Bogan2017} 
and leads to a nontrivial and unavoidable contribution of 
the Dresselhaus SOI to the physics of hole-based devices.
The chief consequence of this interaction is the 
appearance of spin-nonconserving processes
in single-hole tunneling~\cite{Bogan2017} and in coherent control
protocols.~\cite{Bogan2018,Bogan2019-2}

Our previous work~\cite{Bogan2019,Tracy2014,Bogan2017,Bogan2018}
charted only some aspects of the complex physics of holes confined
in the lateral Gated GaAs/AlGaAs double dot under strong SOIs.
This paper is aimed at presenting a comprehensive experimental and
theoretical description of this system, focusing on the two-hole
spin physics.
In the equivalent device confining two electrons, the measurement of
the spin state is carried out electrically via spin-to-charge
conversion.~\cite{Ciorga2000, Ono2002}
This protocol relies on the Pauli spin blockade, i.e.,
the spin-selective tunneling of an electron
from one dot to another conditional on the spin state of
another electron.
This spin-to-charge conversion is routinely used as a readout step
in electron spin coherent control algorithms,~\cite{Nichol2015,Ono2002,Taylor2007,Petta2005,Petta2010,Studenikin2012,Korkusinski2017}
and has also been demonstrated in Si and Ge-based hole devices, 
as well as in the GaAs/AlGaAs double quantum dot confining 
many holes.~\cite{Wang2016,Wang2016-2}
In our system, however, the spin-flip tunneling lifts
the spin blockade.~\cite{Bogan2017}
As a result, we are able to map out the complete energy spectrum 
of the confined two-hole system as a function of the detuning between 
the two dots and the external magnetic field by performing transport
spectroscopy measurements in the large source-drain bias regime.
Moreover, the spin-flip tunneling is a coherent process, 
which, combined with the usual spin-conserving tunneling,
can produce quantum interference between transport channels,
leading to resonant enhancement or suppression of tunneling current.
These unexpected effects are detected experimentally and 
explained quantitatively by a two-site 
Hubbard model, accounting for coherent interdot tunneling 
as well as incoherent coupling to the leads and relaxation processes.
By fitting to the measured transport spectra, we extract the dependencies of 
the system parameters on the magnetic field and gate voltages.
A systematic discussion of the relevant methodological aspects of fitting is presented
to demonstrate that the model provides a precise, quantitative description of the
two-hole double-dot system.
It allows for estimation of key system parameters
(such as, e.g., the spin-flip relaxation time or the effective g-factors),
and therefore can serve as a basis for design of future generations of these devices.
In this discussion we aim to highlight both the strong and the less robust aspects of performing
such detailed fitting.

The paper is organized as follows.
In Sec. II we describe in detail the lateral gated GaAs/AlGaAs
double quantum dot used in this work.
Sec. III contains the description of the theoretical model.
In Sec. IV we present a detailed experimental and theoretical
analysis of the spectra recorded in the blockaded direction,
while in Sec. V we present a similar analysis in the non-blockaded direction.
In Sec. VI we summarize the results of this work.

\section{Double-dot gated hole device}

\begin{figure}[t]
    \includegraphics[width=0.65\textwidth]{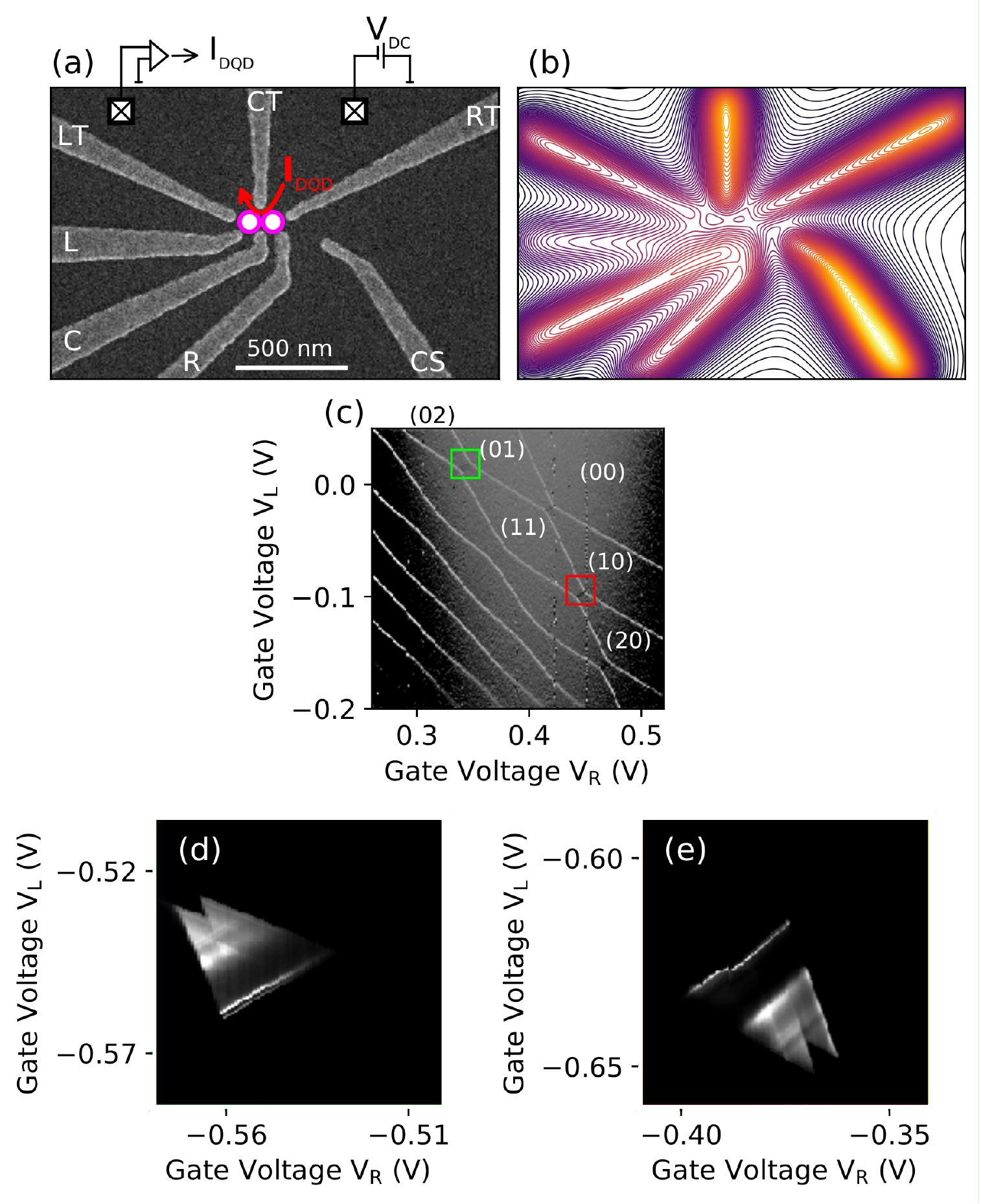}
    \caption{(a) Scanning electron micrograph of the gate layout of the lateral
    double dot confining holes. The transport spectra are recorded by applying the
    source-drain voltage $V_{SD}$ between the hole reservoirs denoted by gray boxes
    and measuring the resulting current.
    (b) A simulated example of the electrostatic potential generated by 
    gate voltages at the hetero-interface confining holes.
    (c) Charge stability diagram of the system measured by the charge sensor created
    by the gate $CS$. The diagram shows addition lines of individual holes 
    admitted into the double 
    dot potential, with the resulting charge configurations indicated in the panel.
    (d) The high-bias transport diagram measured in the high source-drain voltage regime
    ($V_{SD}=2$ mV) close to the resonance of $(11)$ and $(20)$ charge configurations
    denoted in panel (c) by the red rectangle.
    (e) A similar transport diagram for the opposite source-drain bias voltage $V_{SD}=-2$ mV, in the same stability region.
    }
    \label{fig1}
\end{figure}

Figure~\ref{fig1}(a) shows the scanning electron microscope image of 
our lateral double dot device (DQD).\cite{Bogan2019,Tracy2014,Bogan2017,Bogan2018}
The Ti/Au gates are deposited atop the undoped wafer with the
GaAs/ Al$_x$Ga$_{1-x}$As ($x$=50\%) hetero-interface positioned
$65$ nm below the surface.
The two-dimensional hole gas (2DHG) is accumulated at this hetero-interface by 
a global top gate (not shown), separated from the surface gates by 
a 110 nm-thick Al$_2$O$_3$ dielectric layer grown by atomic layer deposition.
The device is cooled down in a dilution refrigerator with the lattice 
temperature below $60$ mK. 
The measured hole temperature was about $100$ mK.

The DQD lateral confinement is created by applying suitable voltages to the surface gates.
Figure~\ref{fig1}(b) shows a calculated electrostatic potential resulting from one such
choice, leading to the formation of a double-dot potential.~\cite{Kyriakidis2002}
The energy detuning between dots is manipulated by the gate voltages 
$V_L$ and $V_R$, applied to the plunger gates $L$ and $R$, respectively,
while the height of the interdot barrier is tuned by adjusting the
voltage on the gate $C$.
Figure~\ref{fig1}(c) shows the addition diagram mapped out as a function of the two
plunger gate voltages by means of the quantum point contact formed by the gate $CS$.
We discuss this charge detection spectroscopy in detail elsewhere.~\cite{Tracy2014,Bogan2017}
The charging diagram shows clear charge addition lines, indicating that
we are able to empty the system completely, generating the charge configuration
$(n_l,n_R)=(00)$, where $n_K$ is the number of holes in dot $K=L,R$.
Further, we add holes one by one to achieve the total occupation of two holes,
realized as charge configurations $(20)$, $(11)$, or $(02)$, as labeled in the diagram.

We probe the energy levels of the two-hole system utilizing the transport
spectroscopy technique in the regime of high source-drain voltage $V_{SD}$, applied 
to the 2DHG leads as shown in Fig.~\ref{fig1}(a).
We measure the resulting tunneling current as a function of the interdot detuning,
the interdot barrier height, and the external magnetic field applied in the direction 
normal to the  plane of the 2DHG.
We focus on tunneling of a hole through the system already confining one "spectator"
particle.
This process can occur in two scenarios, i.e., when the charge configurations $(10)$,
$(11)$, and $(20)$ [Fig.~\ref{fig1}(c), red rectangle], or configurations
$(01)$, $(11)$, and $(02)$ [Fig.~\ref{fig1}(c), green rectangle] are close in energy.
It involves one hole tunneling from the source lead, through each of the dots in sequence,
and out into the drain.
In high-$V_{SD}$ spectroscopy, such tunneling occurs for voltages $(V_L,V_R)$
lying within a transport triangle, shown in Fig.~\ref{fig1}(d) and (e) for the 
first scenario, but with a positive and negative bias voltage, respectively.

\begin{figure}[t]
    \includegraphics[width=0.7\textwidth]{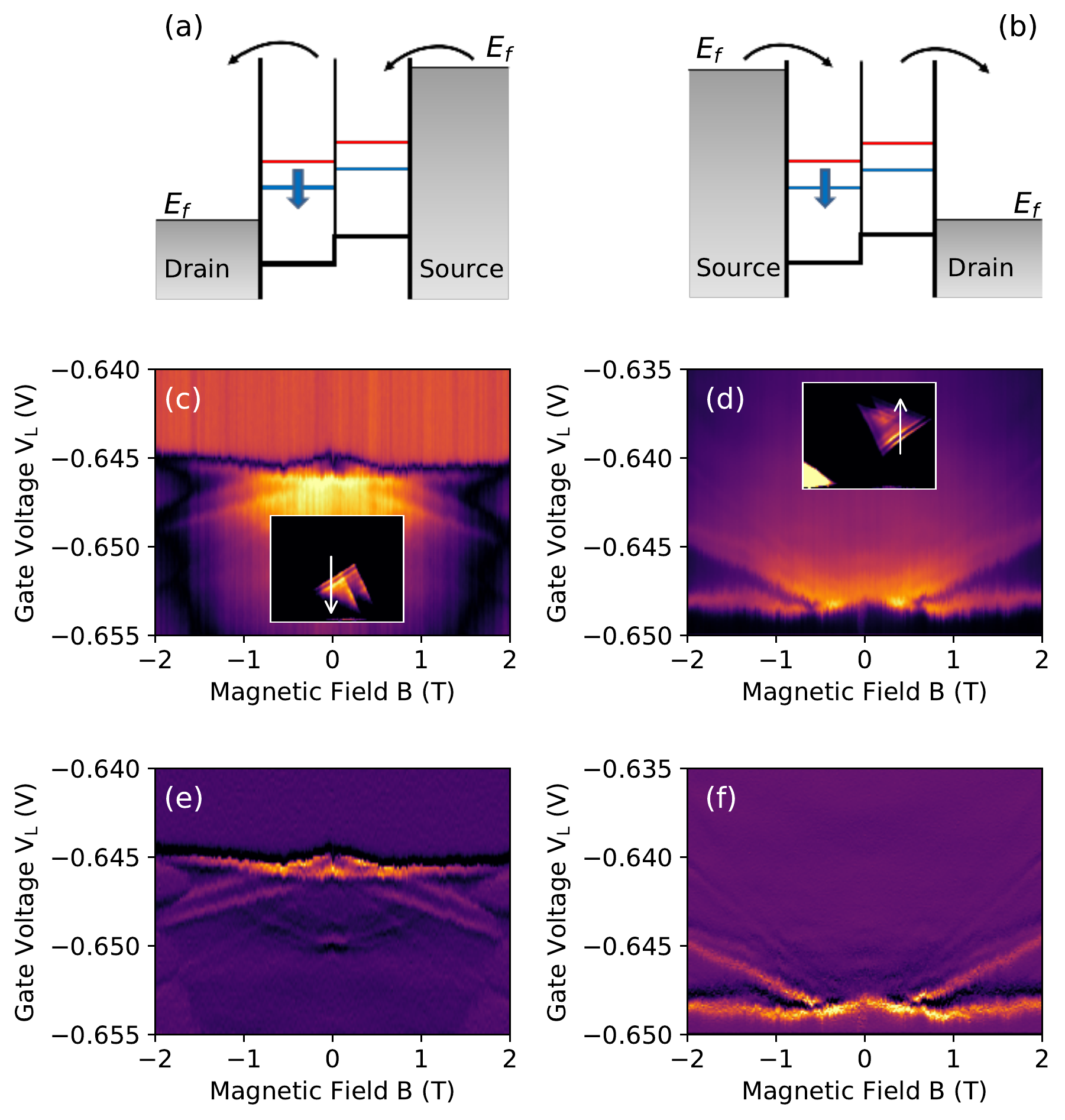}
    \caption{Schematic diagrams of the alignment of single-particle levels of each dot 
    of the system in the "blockade" (a) and "conduction" (b) direction.
    The spin down (up) single-particle levels are denoted in blue (red).
    The solid arrow denotes the "spectator" hole confined in the left dot.
    $E_F$ denotes the Fermi energies in the leads.
    (c) and (d): The tunneling current $I$ resulting from the two alignments, 
    respectively, recorded as a function of the left gate voltage $V_L$ and 
    the magnetic field, with the tunnel barrier defined by the voltage $V_C=-0.16$ $V$. 
    Insets show the corresponding high-bias transport triangles as a function of the voltages $(V_L,V_R)$ for
    one value of the magnetic field, with arrows showing the voltage sweep 
    trajectory followed in the main panels.
    (e) and (f): The current derivative $dI/dV_L$ corresponding to the traces in (c) and 
    (d), respectively, revealing the details of tunneling channels in both directions.}
    \label{fig2}
\end{figure}

We now focus on the transport triangle shown in Fig.~\ref{fig1}(e).
In this segment of the charging diagram there is one hole
confined in the left dot, as presented schematically in Fig.~\ref{fig2}(a) and (b).
These diagrams show the relative alignment of single-particle levels in each dot, with
the blue (red) color denoting the hole spin down (up) level.
The source-drain voltage $V_{SD}=-2$ mV in diagram (a) and 
$V_{SD}=+2$ mV in (b) are applied, resulting in the second hole
tunneling from right to left in the first case, and from
left to right in the second.
The transport triangles corresponding to the two choices of $V_{SD}$, recorded at a finite 
magnetic field, are presented as insets to Fig.~\ref{fig2}(c) and (d), respectively.
Both these triangles are found within the red rectangle in Fig.~\ref{fig1}(c)
and differ only by a choice of interdot detuning which is small on the scale of the full
charging diagram.
Close to the base of each triangle we find several bright lines, corresponding to the
resonant peaks of the tunneling current.
In Fig.~\ref{fig2}(c) and (d) we plot the tunneling current revealing positions of these 
peaks as a function of the gate voltage $V_L$ and the magnetic field, recorded by following 
the trajectory marked in each inset by the white arrow.
The complex set of intersecting lines is visible more clearly in the plots
of the differential current $dI/dV_L$, shown respectively in Fig.~\ref{fig2}(e) and (f).
This is the central experimental result of this work.

In the "forward", or "conduction" direction, corresponding to the source-drain
bias direction as in Fig.~\ref{fig2}(b), the existence of current resonances is expected.
Indeed, a single hole with spin up can tunnel from the
source onto the left-dot spin-up energy level, whereupon either of the holes can tunnel 
to the right dot, and then out into the drain.
On the other hand, in the "blockaded" direction, shown in Fig.~\ref{fig2}(a),
one expects the Pauli spin blockade, occurring where a spin-down hole tunnels from the
source into the right dot.
The resulting triplet $(11)$ configuration cannot convert into a $(20)$ singlet 
configuration without a spin flip.
In our sample, the spin-flip tunneling, leading to the appearance of
the clear tunneling maxima in Fig.~\ref{fig2}(e) (all
diagonal features),
occurs as a consequence of the strong spin-orbit interaction.
We stress that the Pauli blockade prevents the current flow only
in the case of a small interdot detuning, at which
only the singlet-triplet resonance is possible.
For larger detunings, the triplet $(11)$ may become
resonant with a triplet $(20)$ state, enabling tunneling
of carriers without spin flip.
Such a tunneling resonance is seen in Fig.~\ref{fig2}(e)
as the horizontal line, and is a transport feature shared
between hole and electron systems.
Below we provide a systematic, quantitative discussion of all tunneling
maxima appearing in the traces in Fig.~\ref{fig2} and their dependence on
the magnetic field.

\section{Theoretical model}
\subsection{Effective Hamiltonian for holes \label{theory_qual}}

The real-space Hamiltonian for a single heavy hole confined in a quasi-two-dimensional
heterostructure can be written as\cite{Bulaev2005,Bulaev2007,Szumniak2012,Szumniak2013}
\begin{equation}
    \hat{H}_{2D} = \frac{1}{2m_{||}}(\pi_x^2 + \pi_y^2 ) + V_{DDOT}(x,y)
    + \beta ( \pi_-\pi_+\pi_- \sigma_+ + \pi_+\pi_-\pi_+ \sigma_-)
    + \frac{1}{2} g^* \mu_B B_z \sigma_z.
    \label{hamil_eff_rsp}
\end{equation}
Here, $m_{||}$ is the in-plane effective hole mass,
the generalized momentum operator $\bm{\pi}=\bm{p}+\frac{e}{c}\bm{A}$,
where $\bm{p}=-i\hbar\bm{\nabla}$, $\bm{A}$ is the magnetic vector potential, 
$e$ is the electron charge, and $c$ is the speed of light.
The operator $\pi_{\pm} = \pi_x \pm i \pi_y$.
The potential $V_{DDOT}(x,y)$ describes the lateral confinement created by gates,
as seen in Fig.~\ref{fig1}(b).
Also, $g^*$ is the effective Land\'e factor, $\mu_B$ is the (electronic) Bohr magneton, 
$B_z$ is the out-of plane magnetic field, and $\beta$ is the Dresselhaus 
SO parameter.~\cite{Bulaev2005,Bulaev2007}.
Lastly, $\bm{\sigma}$ is the spin operator such that
the projections $\sigma_z = \pm 1$ for the two heavy-hole subbands, respectively, and 
$\sigma_+$ and $\sigma_-$ are respectively the spin raising and lowering operators.
$\hat{H}_{2D}$   can be derived perturbatively from the Luttinger-Kohn
Hamiltonian~\cite{Luttinger1955,Luttinger1956} accounting for the Dresselhaus
SOI.~\cite{Dresselhaus1955,Cardona1988,Bulaev2005,Bulaev2007,Szumniak2012,Szumniak2013}
It can be used to model our gated double-dot device,
for example, within the Heitler-London~\cite{Burkard99,Calderon06} or 
Hund-M\"ulliken~\cite{Hu00,Wiel06,Hatano08} approach.
In these models, quantum molecular energies and molecular orbitals
can be computed based on a realistic double-dot potential profile. 
However, the utilization of $\hat{H}_{2D}$ in modeling the transport 
spectra such as shown in Fig.~\ref{fig2}
is prohibitively expensive computationally owing to costly recomputation of 
$V_{DDOT}$ as gate voltages are adjusted.
Instead, we map the general form (\ref{hamil_eff_rsp}) onto
an effective Hubbard Hamiltonian, whose matrix elements will be fitted to the
experimental data.~\cite{Gimenez2007}
In such approach we lose access to the real-space shape of molecular orbitals,
but are still capable of mapping out the structure of energy levels and
level resonances in a computationally efficient manner.
A detailed comparison of the three models performed with a double-dot system confining
two electrons, is given in Ref.~\onlinecite{Buonacorsi20}.

\begin{figure}[t]
    \includegraphics[width=0.65\textwidth]{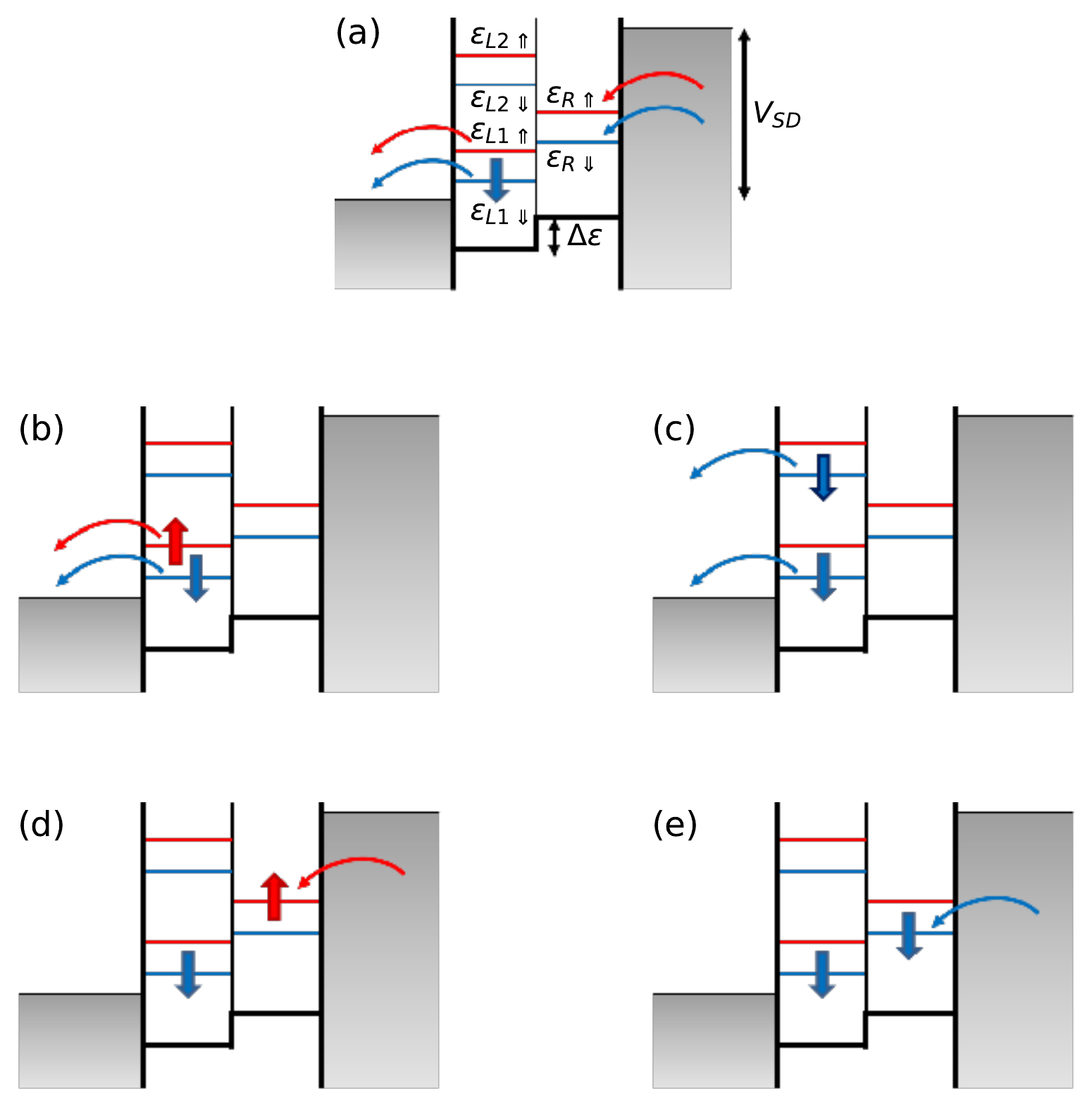}
    \caption{(a) Schematic diagram of the double-dot potential with several
    single-hole states in the left ($L$) and right ($R$) dot. 
    A detuning of $\Delta\varepsilon$ is applied.
    Red (blue) bars represent spin-up (down) states.
    The two-dimensional hole gas in the leads is shown with striped rectangles, 
    with their upper edge denoting the respective Fermi energy of the source (right) 
    and drain (left), separated by the source-drain voltage $V_{SD}$. 
    Tunneling of the hole between the leads and the confined levels
    is denoted by the curved arrows.
    (b) and (c) Two out of four two-hole spin configurations in the charge
    configuration $(20)$ -- the singlet $S(20)$ and the polarized triplet $T_-(20)$.
    Curved arrows denote tunneling channels into the drain considered in this work.
    (d) and (e) Two out of four two-hole spin configurations in the charge configuration
    $(11)$ -- the unpolarized configuration being a part of $S(11)$ and $T_0(11)$ and
    the polarized triplet $T_-(11)$, respectively. Arrows denote the tunneling channels
    from the source into the right dot.}
    \label{fig3}
\end{figure}

We begin by selecting a minimal basis of the single-particle orbitals
relevant for our system.
Figure~\ref{fig3}(a) shows a schematic view of the double-dot potential with 
several dot-resolved orbitals represented by horizontal bars.
Here, the arrow labeled as $\Delta\varepsilon$ is the interdot detuning, while
that marked by $V_{SD}$ represents the source-drain voltage, chosen to induce the 
current flow from right to left.
We enumerate the relevant orbitals in the occupation number representation, with $h^+_{K\sigma}$ ($h_{K\sigma}$) denoting the creation (annihilation) operator of a hole on orbital $K$ 
with spin $\sigma=\Uparrow,\Downarrow$.
In the left dot we consider the levels 
$|L1\Uparrow\rangle = h^+_{L1\Uparrow}|0\rangle$ and
$|L1\Downarrow\rangle = h^+_{L1\Downarrow}|0\rangle$,
which have the same orbital part, but opposite spins
($\sigma = -1$, blue, and $\sigma=+1$, red, respectively).
The second pair, 
$|L2\Uparrow\rangle = h^+_{L2\Uparrow}|0\rangle$ and
$|L2\Downarrow\rangle = h^+_{L2\Downarrow}|0\rangle$,
are the two spin states associated with an excited orbital. 
In the right dot we specify only one pair of levels,
$|R\Uparrow\rangle = h^+_{R\Uparrow}|0\rangle$ and
$|R\Downarrow\rangle = h^+_{R\Downarrow}|0\rangle$, which are the two spin states
associated with the lowest-energy orbital.
Moreover, with curved arrows we indicate the tunneling channels of either spin,
enabling the hole to tunnel from the source into the right dot as well as from the
left dot into the drain.

Utilizing the single-particle basis described above, we model the double-dot system
with the effective Hubbard Hamiltonian of the following form:
\begin{equation}
    \hat{H}_{eff} = \sum_{K}\sum_{\sigma} \varepsilon_{K\sigma} n_{K\sigma}
    + \sum_{K,K'}\sum_{\sigma,\sigma'} t_{K\sigma,K'\sigma'} h^+_{K\sigma}h_{K'\sigma'}
    + \sum_{K,K'}\sum_{\sigma,\sigma'} U_{K\sigma,K'\sigma'} n_{K\sigma}n_{K'\sigma'},
    \label{hamil_Hubbard_generic}
\end{equation}
where $K=L1$, $L2$, $R$, $n_{K\sigma} = h^+_{K\sigma}h_{K\sigma}$ is the density operator
on the orbital $|K\sigma\rangle$, $t_{K\sigma,K'\sigma'}$ are the single-particle hopping
matrix elements, and $U_{K\sigma,K'\sigma'}$ describe the effective Coulomb hole-hole
interactions.

The onsite energies 
$\varepsilon_{K\sigma} = \varepsilon_K(V_L,V_R,B) + \frac{1}{2} g^* \mu_B B \sigma$.
The first term accounts for their dependence on gate voltages and the
magnetic field (via the diamagnetic shift~\cite{Kyriakidis2002}) while
the second term accounts for the Zeeman energies.
Additionally, the energies $\varepsilon_{L1\sigma}$ and $\varepsilon_{L2\sigma}$
can be tuned with respect to $\varepsilon_{R\sigma}$ by the gate voltages.
As shown in Fig.~\ref{fig2}, we tune the voltage $V_L$, which influences both dots
but the left dot more strongly.
As a result, by tuning $V_L$ we tune the detuning between dots
$\Delta\varepsilon = \varepsilon_{L1\Downarrow} - \varepsilon_{R\Downarrow}$.
We convert the gate voltage into the onsite energies using the measured lever arm parameter $\alpha=50$ $\mu$eV/mV, so that 
$\Delta\varepsilon = \alpha V_L$.

The elements $t_{K\sigma,K',\sigma'}$ taken with $\sigma=\sigma'$
represent the spin-conserving tunneling.
No off-diagonal matrix elements connecting $L1$ and
$L2$ states will be generated, as these have been chosen as eigenstates of the left-dot
potential (i.e., $t_{L1\sigma,L2\sigma}=0$).
Of the remaining spin-conserving tunneling elements, we define 
$t_{L1\Downarrow,R\Downarrow} = t_{L1\Uparrow,R\Uparrow} = -t_N$, and
$t_{L2\Downarrow,R\Downarrow}=t_{L2\Uparrow,R\Uparrow}=-t_N'$,
with real an positive $t_N$ and $t_N'$.

Let us now turn to the spin-flip elements.
As evident from the Hamiltonian $\hat{H}_{2D}$, these terms are a consequence of the Dresselhaus SOI, expressed with the generalized momentum in third power.
This operator is of odd symmetry, and therefore
we may expect a strong connection between single-hole states only if these states
are sufficiently different in their orbital part.
Indeed, $t_{L1\Downarrow,L1\Uparrow} = 0$
because these two basis states have the same orbital part.
On the other hand, the element $t_{L1\Downarrow,L2\Uparrow}$
is, in principle, finite.
The impact of a similar intra-dot spin-flip element has been observed in InAs wires
confining electrons.\cite{Pfund2007}
However, in our system we find it to be negligible, as evidenced by two-hole
transport spectra considered in detail below.
The small magnitude of that element was also reported in the theoretical studies of
self-assembled quantum dots confining holes.~\cite{Bulaev2005}
Thus, we consider the spin-flip matrix elements of the form
$t_{L1\Uparrow,R\Downarrow}=t_{L1\Downarrow,R\Uparrow}=-it_F$ and
$t_{L2\Uparrow,R\Downarrow}=t_{L2\Downarrow,R\Uparrow}=-it_F'$, with
real and positive $t_F$ and $t_F'$.

The difference in the microscopic origin of $t_N$ and $t_F$ leads to their
different dependence on the magnetic field.
In our earlier studies, we have found that $t_N$ decreases with the increase
of $B$ owing to the diamagnetic squeezing of the single-dot states.
On the other hand, the $B$-field dependence of $t_F$ was shown to be nonmonotonic
in systems with weak tunnel coupling (i.e., larger dots),~\cite{Bogan2017} 
but exhibited a monotonic exponential decay in strongly coupled systems 
(smaller dots).~\cite{Bogan2019-2}

In our previous work,~\cite{Bogan2018} we have extensively studied the single-hole system
defined by the Hamiltonian (\ref{hamil_Hubbard_generic})
utilizing the Landau-Zener-St\"uckelberg as well as the photon-assisted tunneling
spectroscopy and have shown that this model is sufficient to describe the essential physics
of our double dot confining a single hole.
Presently we will build upon this model to treat the two-hole system.

\subsection{Two-hole Hamiltonian \label{two_hole_qualitative}}

We consider only those two-hole configurations in which
one hole is placed in the left dot.
Figures~\ref{fig3}(b)--(e) show several such configurations.
We begin with the charge configuration $(20)$, in which both holes occupy
the left dot.
The singlet configuration
$|S(20)\rangle = h_{L1,\Uparrow}^+h_{L1,\Downarrow}^+|0\rangle$
is shown in Fig.~\ref{fig3}(b).
This is the only configuration in which both holes occupy the orbital $L1$
(with opposite spins).
Owing to the Pauli principle, both orbitals $L1$ and $L2$ are needed to construct the
doubly-occupied triplets,
$|T_-(20)\rangle =
h_{L1,\Downarrow}^+h_{L2,\Downarrow}^+|0\rangle$,
$|T_0(20)\rangle = \frac{1}{\sqrt{2}}\left( 
h_{L1,\Uparrow}^+h_{L2,\Downarrow}^+ -
h_{L2,\Uparrow}^+h_{L1,\Downarrow}^+\right) |0\rangle$,
and
$|T_+(20)\rangle = h_{L1,\Uparrow}^+h_{L2,\Uparrow}^+|0\rangle$.
The polarized triplet $|T_-(20)\rangle$ is shown schematically in Fig.~\ref{fig3}(c).

In the charge configuration $(11)$, we take the
singlet $|S(11)\rangle = \frac{1}{\sqrt{2}}\left( 
h_{L1,\Uparrow}^+h_{R,\Downarrow}^+ + 
h_{R,\Uparrow}^+h_{L1,\Downarrow}^+\right) |0\rangle$,
shown schematically in Fig.~\ref{fig3}(d).
Note that the diagram represents only the second configuration in this superposition.
The unpolarized triplet
$|T_0(11)\rangle = \frac{1}{\sqrt{2}}\left( 
h_{L1,\Uparrow}^+h_{R,\Downarrow}^+ -
h_{R,\Uparrow}^+h_{L1,\Downarrow}^+\right) |0\rangle$
is similar to the singlet except for the phase factor.
Finally, the two polarized triplets are
$|T_-(11)\rangle = h_{L1,\Downarrow}^+h_{R,\Downarrow}^+|0\rangle$ and
$|T_+(11)\rangle = h_{L1,\Uparrow}^+h_{R,\Uparrow}^+|0\rangle$,
of which the former is shown in Fig.~\ref{fig3}(e).

We order the two-hole configurations as \{$|S(20)\rangle$, $|T_-(20)\rangle$,
$|T_0(20)\rangle$, $|T_+(20)\rangle$, $|T_-(11)\rangle$, $|S(11)\rangle$, 
$|T_0(11)\rangle$, $|T_+(11)\rangle$\}.
In this basis we arrive at the following two-hole Hamiltonian matrix:
\begin{equation}
    \hat{H}_{2H} = \left[
    \begin{array}{cccccccc}
    E_{S(20)} & 0 & 0 & 0 & it_F & -\sqrt{2}t_N & 0 & -it_F\\
    0 & E_{T(20)} - E_Z & 0 & 0 & -t_N' & \frac{i}{\sqrt{2}}t_F'
    & -\frac{i}{\sqrt{2}}t_F' & 0 \\
    0 & 0 & E_{T(20)} & 0 & -\frac{i}{\sqrt{2}}t_F' & 0 &
    -t_N' & -\frac{i}{\sqrt{2}}t_F'\\
    0 & 0 & 0 & E_{T(20)}+E_Z & 0 & -\frac{i}{\sqrt{2}}t_F' &
    -\frac{i}{\sqrt{2}}t_F' & -t_N' \\
    -it_F & -t_N' & \frac{i}{\sqrt{2}}t_F' & 0 & E_{T(11)} - E_Z 
    & 0 & 0 & 0 \\
    -\sqrt{2}t_N & -\frac{i}{\sqrt{2}}t_F' & 0 & 
    \frac{i}{\sqrt{2}}t_F' & 0 & E_{S(11)} & 0 & 0 \\
    0 & \frac{i}{\sqrt{2}}t_F' & -t_N' & 
    \frac{i}{\sqrt{2}}t_F' & 0 & 0 & E_{T(11)} & 0 \\
    it_F & 0 & \frac{i}{\sqrt{2}}t_F' & -t_N' & 0 & 0 & 0 &
    E_{T(11)}+E_Z
    \end{array}
    \right].
    \label{th_hamil}
\end{equation}
Here, $E_{S(20)}$ is the energy of the doubly-occupied singlet configuration
$|S(20)\rangle$, while $E_{T(20)}$ is that of the unpolarized triplet
$|T_0(20)\rangle$.
In terms of the elements of the Hubbard Hamiltonian (\ref{hamil_Hubbard_generic}),
these energies can be expressed as
$E_{S(20)} = 2\varepsilon_{L1} +U_{L1\Uparrow,L1\Downarrow}$ 
and
$E_{T(20)} = \varepsilon_{L1} + \varepsilon_{L2}  +U_{L1\Uparrow,L2\Downarrow}$.
Here, $\varepsilon_K = \frac{1}{2}(\varepsilon_{K\Downarrow} + \varepsilon_{K\Uparrow})$
is the orbital part of the single-hole energy.
The Coulomb interaction terms $U$ will be obtained
by fitting to the experimental data.
Here let us only mention that
the interaction term appearing in the triplet energy can be decomposed as
$U_{L1\Uparrow,L2\Downarrow} = U_{L1\Uparrow,L2\Downarrow}^D +
U_{L1\Uparrow,L2\Downarrow}^X$, i.e., the direct Coulomb repulsion and the exchange terms,
respectively.
As can be seen, the energy gap $\Delta E_{ST} = E_{T(20)} - E_{S(20)}
= \varepsilon_{L2} - \varepsilon_{L1} +U_{L1\Uparrow,L2\Downarrow} - U_{L1\Uparrow,L1\Downarrow}$ 
appears primarily due to the fact that
the triplet configuration is created by placing one of the holes on an excited
single-particle orbital $L2$ in the left dot, while in the singlet configuration
both holes occupy the lowest single-particle orbital $L1$.
The interaction terms also contribute to this gap.
The energies of the two polarized triplets, $|T_-(20)\rangle$ and $|T_+(20)\rangle$, 
are respectively lower and higher than that of the unpolarized triplet by the 
Zeeman energy $E_Z = g^*\mu_B B$.
We stress that the singlet triplet gap $\Delta E_{ST}$ depends on the magnetic field.
Moreover, there exists a magnetic field $B_{ST}$, for which
we find the singlet-triplet transition in the ground state of the
doubly-occupied dot, from the singlet to the polarized triplet
$|T_-(20)\rangle$.~\cite{Kyriakidis2002,Ciorga2002}

In the $(11)$ charge configuration,
the singlet and unpolarized triplet energies are respectively
$E_{S(11)} = \varepsilon_{L1} + \varepsilon_{R}
+U^S_{L1\Uparrow,R\Downarrow}$
and
$E_{T(11)} = \varepsilon_{L1} + \varepsilon_{R} +U^T_{L1\Uparrow,R\Downarrow}$.
Assuming equal $g^*$ factors in each dot, these energies differ only by the 
interaction terms $U$, and specifically by the
different contributions of the hole-hole Coulomb exchange.
However, in our system the resultant energy gap between the states $|S(11)\rangle$ 
and $|T_0(11)\rangle$ is negligibly small, and in the following we treat these two
configurations as degenerate in energy, i.e., $E_{S(11)}=E_{T(11)}$, and assume
$U^S_{L1\Uparrow,R\Downarrow}=U^T_{L1\Uparrow,R\Downarrow}=0$.
For the charge configuration $(11)$ there are no level crossings, and the
polarized triplet $|T_-(11)\rangle$ is the lowest-energy configuration for any
finite magnetic field owing to the Zeeman energy $E_Z$.

The next step in our computational procedure is to diagonalize the above two-hole Hamiltonian
to obtain the quantum-molecular two-hole states.
We do this for each value of the magnetic field and each gate voltage $V_L$.
As a result, we obtain eight double-dot, two-hole quantum molecular states $|i\rangle$,
with $i=1,\dots,8$, expressed, in general, as linear combinations of all eight
two-hole basis states:
\begin{equation}
    |i\rangle = \sum_{j=1}^8 A_j^{(i)} |j\rangle,
    \label{hybrid-2h}
\end{equation}
where the index $j$ enumerates the basis states in the order given above.

\subsection{Calculation of the tunneling current}

To calculate the tunneling current flow we utilize the density matrix formalism.
We define the density matrix $\hat{\varrho}$ in the basis of the eight
two-hole quantum molecular states plus two states for the single "spectator" hole, 
$|L1\Uparrow\rangle$ and $|L1\Downarrow\rangle$.
The dynamics of the system can be calculated by solving the usual master equation
\begin{equation}
    \frac{d}{dt} \hat{\varrho}(t) = i[\hat{\varrho},\hat{H}] 
    + \hat{\Gamma}^{(in)}\hat{\varrho} + \hat{\Gamma}^{(out)}\hat{\varrho}
    + \hat{\Gamma}^{(SF)}\hat{\varrho},
\end{equation}
where the second, third, and fourth term on the right-hand side describe the 
hole tunneling in, out, and spin-flip relaxation processes, respectively.
In the molecular basis $\{|i\rangle\}$ the Hamiltonian is diagonal, which sets 
the commutator to zero allowing us to track exclusively
the time evolution of the diagonal 
density matrix elements $n_i$ (i.e., level occupations).
The long-time (steady-state)
behaviour of our system is obtained by setting $\frac{d}{dt} \hat{\varrho}(t)=0$.
As a result, we only account for the incoherent processes transferring the hole from
and to the leads ($\hat{\Gamma}^{(in)}$ and $\hat{\Gamma}^{(out)}$ respectively)
and the spin-flip relaxation processes $\hat{\Gamma}^{(SF)}$.
Details of the treatment of the former two processes are given in Appendix~\ref{app:leads},
while the latter is described in Appendix~\ref{app:relaxation}.
The system of ten algebraic equations to be solved as a function of gate voltages
and the magnetic field is presented in Appendix~\ref{app:dmf}.
Once the occupations of the two-hole molecular levels are obtained,
the current is computed as the total charge flux out of the double dot:
\begin{equation}
    I = e\sum_{i=1}^8 \sum_{\sigma_1}\sum_{\sigma_2}
    \Gamma_i^{(out)}(\sigma_1,\sigma_2) n_i.
    \label{tot_current}
\end{equation}

\section{Transport spectra of two holes in the blockaded direction}

Our theoretical description is formulated for the "blockaded" direction of transport,
i.e., one in which the sequence of charge configurations is $(10)\rightarrow$ $(11)\rightarrow$
$(20)\rightarrow$ $(10)$.
Here, the spectator hole is confined in the left dot.
The transport spectra as a function of the magnetic field 
recorded at this configuration are shown in Fig.~\ref{fig2}(c) and (e)
for the voltage region marked in Fig.~\ref{fig1}(c) by the red square.
In that case, the voltage applied to the gate $C$ was $V_C=-0.16$ $V$.
In this Section we will reproduce these spectra with our theoretical model.

\subsection{Positions of current maxima \label{idealized}}

\begin{figure}[t]
    \includegraphics[width=0.7\textwidth]{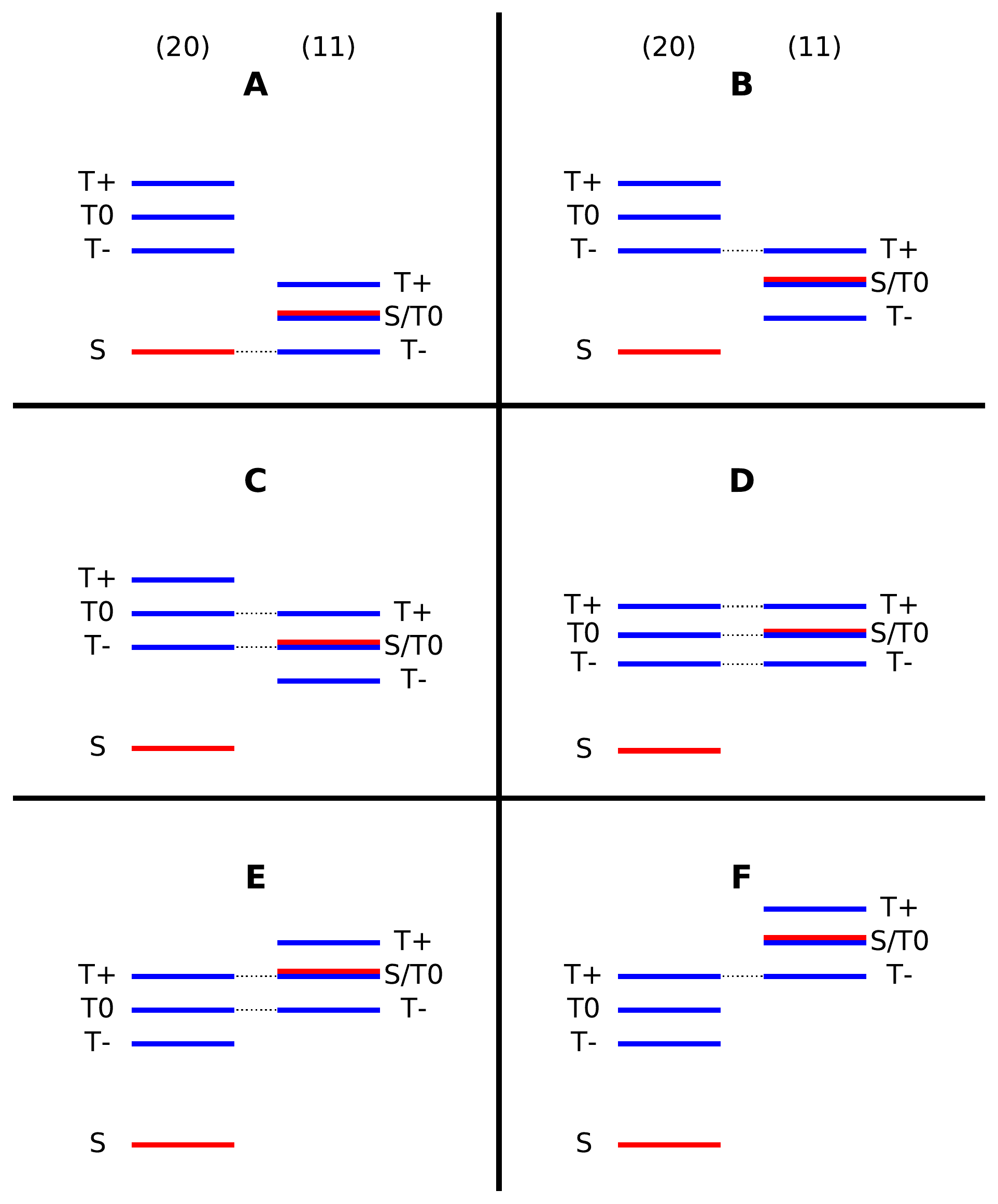}
    \caption{Different realizations of tunneling resonances between
    $(20)$ (left in each panel) and $(11)$ (right) charge configurations relevant 
    for the blockaded  direction. 
    Panels A through F show level alignments with increasing negative detuning.
    Red (blue) color corresponds to singlet (triplet) levels.}
    \label{fig5}
\end{figure}

Formation of conduction channels across the double dot 
occurs when any $(11)$ configuration is close to resonance with a 
$(20)$ configuration, the former being connected to the source and the latter 
-- to the drain.
In Fig.~\ref{fig5} we show all possible resonant level alignments relevant for
the blockaded direction.
We choose to represent these alignments schematically for the magnetic field
$B<B_{ST}$, i.e., when the lowest-energy $(20)$ configuration is the singlet
$S(20)$, but all these resonances can be traced to higher fields with appropriate
changes in their ordering.
Further, assuming a short spin-flip relaxation time $T_{SF}$,
the tunneling will be mediated predominantly by the $|T_-(11)\rangle$ configuration.
First we consider the $|T_-(11)\rangle$ -- $|S(20)\rangle$ resonance
shown in Fig.~\ref{fig5}A.
Transport maximum will be recorded if $E_{S(20)}=E_{T-(11)}$.
With the detuning $\Delta\varepsilon = \varepsilon_{L}-\varepsilon_R$, 
the above condition is fulfilled for $\Delta\varepsilon_{A} =  
-U_{L1\Uparrow,L1\Downarrow} - E_Z$.
The doubly-occupied configuration has to be detuned to compensate
for the charging energy, while the Zeeman contribution results from
the different spin polarization of the two configurations.
As a result, we expect a current maximum for a negative detuning which shifts 
approximately linearly with the magnetic field, with some nonlinearity arising
from the dependence of the charging energy on $B$.
A similar analysis for the 
$|T_-(11)\rangle$ -- $|T_-(20)\rangle$ resonance (Fig.~\ref{fig5}D)
reveals the detuning
$\Delta\varepsilon_{D} =  -U_{L1\Uparrow,L1\Downarrow} - \Delta E_{ST}$.
Relative to the detuning $\Delta\varepsilon_{T-,S}$, this position will be shifted 
to more negative values by the singlet-triplet gap $\Delta E_{ST}$,
but will not be modified by the Zeeman term.
The two remaining resonances, $|T_-(11)\rangle$ -- $|T_0(20)\rangle$
(Fig.~\ref{fig5}E) and $|T_-(11)\rangle$ -- $|T_+(20)\rangle$ (Fig.~\ref{fig5}F),
are expected respectively at $\Delta\varepsilon_{E} =
\Delta\varepsilon_{D} - E_Z$ and 
$\Delta\varepsilon_{F} = \Delta\varepsilon_{D} - 2E_Z$.
They are shifted from the resonance D by respectively once and twice
Zeeman energy, accounting for the change of spin polarization by one and two.
In the case of fast spin relaxation we expect that only the $|T_-(11)\rangle$ state
will be occupied, and the transport spectra
will reveal the energy structure of the doubly-occupied charge configuration $(20)$.

\begin{figure}[t]
    \includegraphics[width=0.7\textwidth]{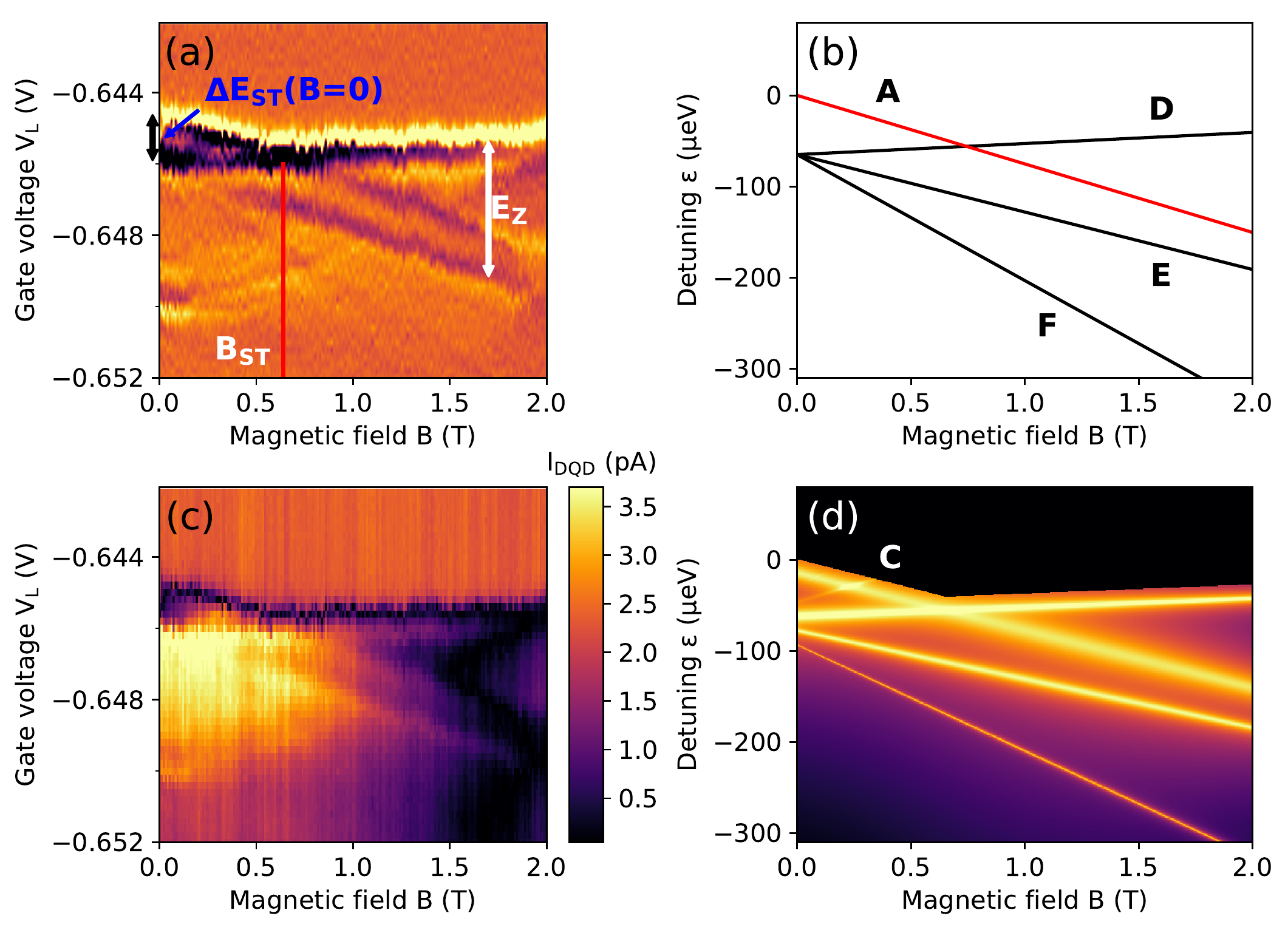}
    \caption{(a) Differential current $dI/dV_L$ as a function of the
    gate voltage $V_L$ and the magnetic field at conditions identical to
    Fig.~\ref{fig2}(e). Note the identification of characteristic elements
    of the spectrum of $(20)$ charge configuration.
    (b) Calculated positions of the transport maxima as a function of detuning
    and the magnetic field. 
    Labels correspond to the resonances shown in Fig.~\ref{fig5}.
    Panels (c) and (d) show respectively the tunneling current $I$ 
    measured as a 
    function of the gate voltage $V_L$ and the magnetic field at conditions 
    identical to Fig.~\ref{fig2}(c), and the current calculated theoretically.}
    \label{fig6}
\end{figure}

To connect our description to the experimental data, we focus on the dataset
presented earlier in Fig.~\ref{fig2}(e), shown
in Fig.~\ref{fig6}(a) in higher resolution.
We find that, as expected, at zero magnetic field
we see only two current maxima as a function of
the voltage $V_L$, separated by $\Delta V_L = 1.3$ mV.
Using the lever arm $\alpha = 50$ $\mu$eV$/$mV, we estimate this gap
to be $65$ $\mu$eV.
This is the $(20)$ singlet-triplet gap $\Delta E_{ST}(B=0)$.
Next, we extract the effective $g$-factor $g^*=1.35$ by measuring the detuning
corresponding to the Zeeman energy as marked in Fig.~\ref{fig6}(a).
Finally, we observe a crossing of two transport maxima at the magnetic field 
$B_{ST}=0.72$ T.
This is the signature of the singlet-triplet transition in the $(20)$ configuration.
At this magnetic field we have
$E_{ST}(B=B_{ST}) = g^*\mu_B B_{ST}$, as the Zeeman energy is equal
to the singlet-triplet gap.
From that equation we obtain $\Delta E_{ST}(B=B_{ST}) = 56.25$ $\mu$eV
and, assuming a linear dependence of $\Delta E_{ST}$ on the field,
we obtain $\Delta E_{ST}(B) = ( 65.00 - 12.15 B) $ $\mu$eV.
The only parameter left unaccounted for in our analysis is the charging
energy $U_{L1\Uparrow,L1\Downarrow}$.
Since this energy renormalizes all resonance detunings equally, we are unable to
extract it from the experimental data and we will treat it as a reference energy, i.e.,
we will calibrate our detuning as $\Delta\varepsilon_A(B=0) = 0$.
Figure~\ref{fig6}(b) shows the position of transport resonances as a function
of detuning and the magnetic field calculated with our model.
We recover a one-to-one correspondence with the experimental result,
including all characteristic points of the spectrum.

Next we reproduce theoretically the amplitudes of the transport peaks.
Tracing the resonances A, D, and E from Fig.~\ref{fig5}, we find that 
the tunnel couplings between the relevant $(20)$ and $(11)$ configurations
are set up in the Hamiltonian (\ref{th_hamil}) by tunneling elements 
$t_F$, $t_N'$, and $t_F'$.
We also find that the resonance F is not provided for in that matrix,
as the relevant off-diagonal matrix element is zero.
This is due to the fact that such a coupling would require a double spin flip
(from configuration $T_-(11)$ to $T_+(20)$) which is not accounted for in our
simple Hamiltonian.
Nonetheless, in the experimental data we find a very weak resonant peak
at the line F, originating most likely from a higher-order tunneling process,
and to enable it in our model we introduce an additional effective matrix element 
$\tau$.

We have extracted the magnetic field evolution of the elements
$t_N$ and $t_F$ in our earlier work in the single-hole regime.~\cite{Bogan2018}
By fitting to the experimental data, we obtained
$t_N(B) = t_N^{(0)}\exp(-B^2/2B_N^2)$ with $t_N^{(0)}=0.24$ $\mu$eV
and $B_N = 1.33$ T for the fundamental spin-conserving element,
and $t_F(B) = (t_F^{(0)} + t_F^{(2)}B^2) \exp(-B^2/B_F^2)$
with $t_F^{(0)} = 0.1$ $\mu$eV, $t_F^{(2)} = 0.27$ $\mu$eV $/$T,
and $B_F = 1.28$ T.
We were unable to establish the parameters $t_N'(B)$ and $t_F'(B)$
in an independent fashion, but expect that they will exhibit a similar
general behavior as the fundamental ones and be somewhat smaller.
For the present discussion we take model values
$t_N'(B) = 0.75 t_N(B)$ and $t_F'(B) = 0.75 t_F(B)$.
Moreover, the effective double-flip tunneling element is taken as 
$\tau = t_F'(B)/10$.

To complete our theoretical model, we also require the estimate of
the spin-flip relaxation time $T_{SF}$ and the rates $\Gamma^{(in)}$ and
$\Gamma^{(out)}$ characterizing the tunnel coupling to the leads.
The spin-flip relaxation rate was measured by us in Ref.~\onlinecite{Bogan2019}
where we found $T_{SF}(B) = T_{SF}^{(0)}B^{-5}$ with 
$T_{SF}^{(0)} = 2.5$ $\mu$s T$^5$.
However, we find that this functional relationship was obtained for a single hole
in a single lateral gated dot and in relatively high magnetic fields and does not
appear to be appropriate in our low-field two-hole system.
Indeed, if we assume such long spin-flip times in our simulation,
the contribution from excited states features much more prominently
than is actually seen in experiment, as further discussed below.
We find that our simulations satisfactorily correspond to the experimental data 
with a field-independent $T_{SF}=500$ ns.
This large discrepancy is due most likely to the fact that
$T_{SF}$ is strongly renormalized by the cotunneling effects brought about by
the connection to the leads.~\cite{Qassemi2009}
This effect did not occur in our earlier measurements,~\cite{Bogan2019}
in which the quantum dot states were kept away from the conduction window,
which prevented the cotunneling from taking place.
Further, we take $\Gamma^{(in)}=\Gamma^{(out)}=2$ GHz, 
consistent with our previous work.~\cite{Bogan2018}

Results of our calculations are presented in Fig.~\ref{fig6}(d) and
compared with the measured current shown in Fig.~\ref{fig6}(c).
The experimental result is a zoomed-in version of the dataset from
Fig.~\ref{fig2}(c).
In the theoretical plot, the region appearing in black in the upper part of
the image corresponds to the energy blockade, i.e., the alignment of levels
in which all energies of the $(20)$ configuration lie above the lowest-energy
level of the $(11)$ configuration.
In this case, the system would be blockaded in the $|T_-(11)\rangle$ state
leading to the collapse of the tunneling current.
Our steady-state approximation describes this accumulation of charge only partially,
giving a gradual decrease of the current as the detuning is made more positive.
To compensate for this, the energy blockade was introduced manually by removing
the current in the blockaded regions, which
results in a better agreement with experimental data.
Further, our calculation reproduces all features of the experimental data 
with one additional maximum, denoted by the label C.
This maximum corresponds to the alignment of levels as in the panel C
of Fig.~\ref{fig5}, in which an excited state $T_0(11)$ or $S(11)$ 
is in resonance with the state $T_-(20)$, and a similar resonance appears
between $T_+(11)$ and $T_0(20)$.
On the other hand, the ground state $T_-(11)$ is not aligned with any
states in the $(20)$ configuration.
In such case, with a sufficiently short spin relaxation time $T_{SF}$
we would again experience an energy blockade, with charge accumulation on
the level $T_-(11)$.
A residual tunneling current appears only because our value of $T_{SF}$
is still slightly too large compared with the experimental data, in which the feature 
C is not well resolved.
This enables the two resonances of excited states to mediate the tunneling
before the system can relax to the lowest level.
We study this additional maximum in our next dataset.
We note that a similar nontrivial maximum should appear in the alignment of levels
labeled B in Fig.~\ref{fig5}.
We do not appear to detect it, as in addition to the relaxation argument
the tunneling in this alignment would require a double spin-flip, and therefore
would be characterized by the very small element $\tau$.

\subsection{Tunneling amplitudes}

\begin{figure}[t]
    \includegraphics[width=0.7\textwidth]{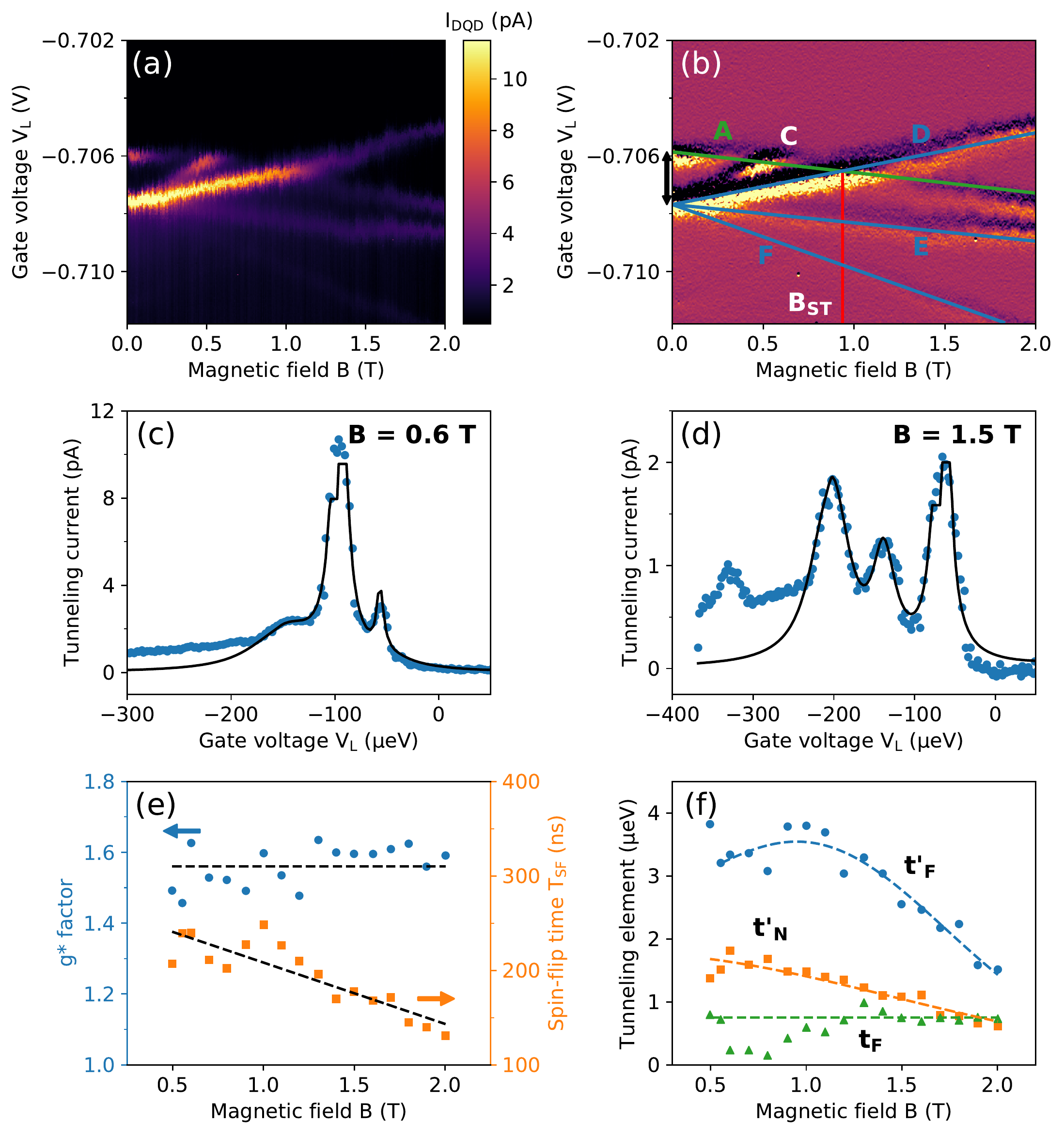}
    \caption{(a) Tunneling current and (b) the derivative $dI/dV_L$ 
    as a function of the gate voltage
    $V_L$ and the magnetic field with the interdot gate voltage
    $V_C=-0.11$ $V$. 
    Note the identification (letter labels) of characteristic elements of 
    the spectrum of $(20)$ charge configuration.
    The black double arrow in (b) marks the zero-field
    singlet-triplet gap $\Delta E_{ST}(B=0)$.
    Panels (c) and (d): the tunneling current as a function of the
    left gate voltage $V_L$ extracted from (a) at the magnetic field $B=0.6$ T
    and $1.5$ T, respectively; experimental data (symbols) and theoretical fits (lines).
    (e) Effective $g^*$ factor (blue) and the spin-flip relaxation time $T_{SF}$
    (orange) as a function of the magnetic field extracted by fitting the theoretical model.
    (f) Fitted values of the spin-conserving ($t_N'$, orange) and spin-flipping
    ($t_F$, green, and $t_F'$, blue) matrix elements as a function of the magnetic field.
    Dashed lines are fits of model theoretical curves (see text).
    }    
    \label{fig7}
\end{figure}

In this Section, we will analyze quantitatively
selected tunneling elements and the tunnel coupling to the leads.
In Fig.~\ref{fig7}(a) and (b) we show the result of another transport 
spectroscopy measurement, this time with a slightly higher interdot barrier
(smaller coupling), set by the voltage $V_C=-0.11$ $V$.
We fit the theoretical transport curves to the experimental data
at several values of the magnetic field.
For example, in Fig.~\ref{fig7}(c) we show the measured (symbols) and fitted
current (line) at $B=0.6$ T, while a similar fit at $B=1.5$ T is shown in
Fig.~\ref{fig7}(d).
The fitting is performed with the assumption of the couplings with
both leads of $\Gamma = 2$ GHz, however all other model parameters
are treated as variables in the procedure.
Overall, we achieve a very close fit with the exception of the region
of large negative detunings [left-hand side of the two panels, or the
lower region of Fig.~\ref{fig7}(a)].
This discrepancy between the theory and the experiment can be traced to the
incoherent off-resonant leakage current across the device, which is
not taken into account in the model.
We also note the signature of the resonance F in the left-hand side of panel (d),
which is not reflected in the theoretical curve as we set the double spin-flip
element $\tau = 0$.

An important methodological remark concerns the treatment of the maxima D
corresponding to the highest peaks in Fig.~\ref{fig7}(c) and (d).
Our theoretical model assumes explicitely the formation of fully coherent
two-hole molecular states of the form of Eq.~(\ref{hybrid-2h}) whenever
different two-hole configurations are in resonance. 
As evident from Fig.~\ref{fig5}, such resonances occur for each of the maxima
A, D, and E.
However, the alignment D is special in that the resonance occurs for all
available states of the $(11)$ charge configurations.
This resonance, therefore, is not affected by the spin relaxation process,
and will result in a very strong current maximum.
In reality, this resonance is affected by the sample environment, and
in particular by the charge noise.
This charge noise was shown to influence significantly the
transport spectra in Landau-Zener-St\"uckelberg-Majorana and the
photon-assisted tunneling experiments.~\cite{Bogan2018}
The charge oscillators present in the vicinity
of our system will introduce a randomly fluctuating
effective detuning between different configurations,
thereby disrupting the resonance condition 
and decreasing the effective current.
These effects are not taken into account in our model, therefore, in order to achieve correspondence between the theoretical model and the
experimental data, we truncated the peaks D, as clearly seen in
Fig.~\ref{fig7}(c) and to a lesser extent in panel (d), attempting
to fit to the width of the peak D rather than to its height.
Without this truncation, the theoretical values of the current close to the resonance
appear by a factor of about $5$ too large.
A similar issue does not occur for the maxima A and E, since in these alignments
at least one level of the $(11)$ configuration is not on resonance and the tunneling
process is naturally attenuated by the spin-flip relaxation with the
characteristic time $T_{SF}$.
This is why both the width and the height of these maxima are well reproduced
by the theory.
Note that the peak D in our model is characterized
by the spin-conserving tunneling element,
while the peaks A and E -- by the spin-flip
tunneling elements.

In Fig.~\ref{fig7}(e) and (f) we show the parameters of the model extracted
in the fitting procedure as a function of the magnetic field.
In panel (e), we find an approximately constant value of the effective $g$-factor
$g^*\approx 1.56$ (blue), while the spin-flip relaxation time 
decreases with the increase of the field as
$T_{SF} = T_{SF}^{(0)} + T_{SF}^{(1)} B$, with $T_{SF}(0)=273.21$ ns
and $T_{SF}^{(1)}=-64.98$ ns$/$T (orange).
As we already saw in Fig.~\ref{fig6}, the extracted value of $T_{SF}$ is smaller
than that measured by us in an isolated dot,~\cite{Bogan2019} particularly
at lower magnetic fields, and exhibits a different functional dependence on the  
field (linear as opposed to $B^{-5}$).
In panel (f) we present the extracted tunneling matrix elements.
The spin-conserving element $t_N'$ (orange) is extracted from the maximum D,
the spin-flipping element $t_F$ (green) - from the maximum A, and
the spin-flipping element $t_F'$ (blue) - from the maximum E.
We do not fit the spin-conserving element $t_N$, as in this alignment
there is no current maximum governed by it at nonzero fields; we will 
come back to this element later on.
In general, we find the expected general decrease of all values with the
increase of the magnetic field, with the exception of $t_F$ (green), which
appears to be lowest for lower values of $B$.
However, in the region of $B$ between $0.6$ T and $1.1$ T the peak A
is obscured by the maximum D, making a reliable fitting very difficult.
We expect that the nonmonotonic behavior of $t_F$ is therefore
an artifact and approximate a field-independent value of $t_F = 0.75$ $\mu$eV
(dashed green line).
The model fits to $t_N'$ and $t_F'$ (orange and blue dashed lines, respectively)
are postulated in analogy to the relationships taken in the previous Section.
We find a very good approximation of both progressions with the following
model formulas:
$t_N' = t_N'^{(0)}\exp(-B^2/2B_N'^2)$, with $t_N'^{(0)} = 1.78$ $\mu$eV and
$B_N' = 1.45$ T for the spin-conserving element, and
$t_F' = (t_F'^{(0)} + t_F'^{(2)}B^2) \exp(-B^2/B_F'^2)$
with $t_F'^{(0)} = 2.64$ $\mu$eV,
$t_F'^{(2)} = 4.12$ $\mu$eV$/$T$^2$, and
$B_F' = 1.24$ T for the spin-flip element.
We note that the overlap of maxima D and E at magnetic fields lower than
$0.5$ T prevents us from extracting reliable values for all model parameters.

\begin{figure}[t]
    \includegraphics[width=0.7\textwidth]{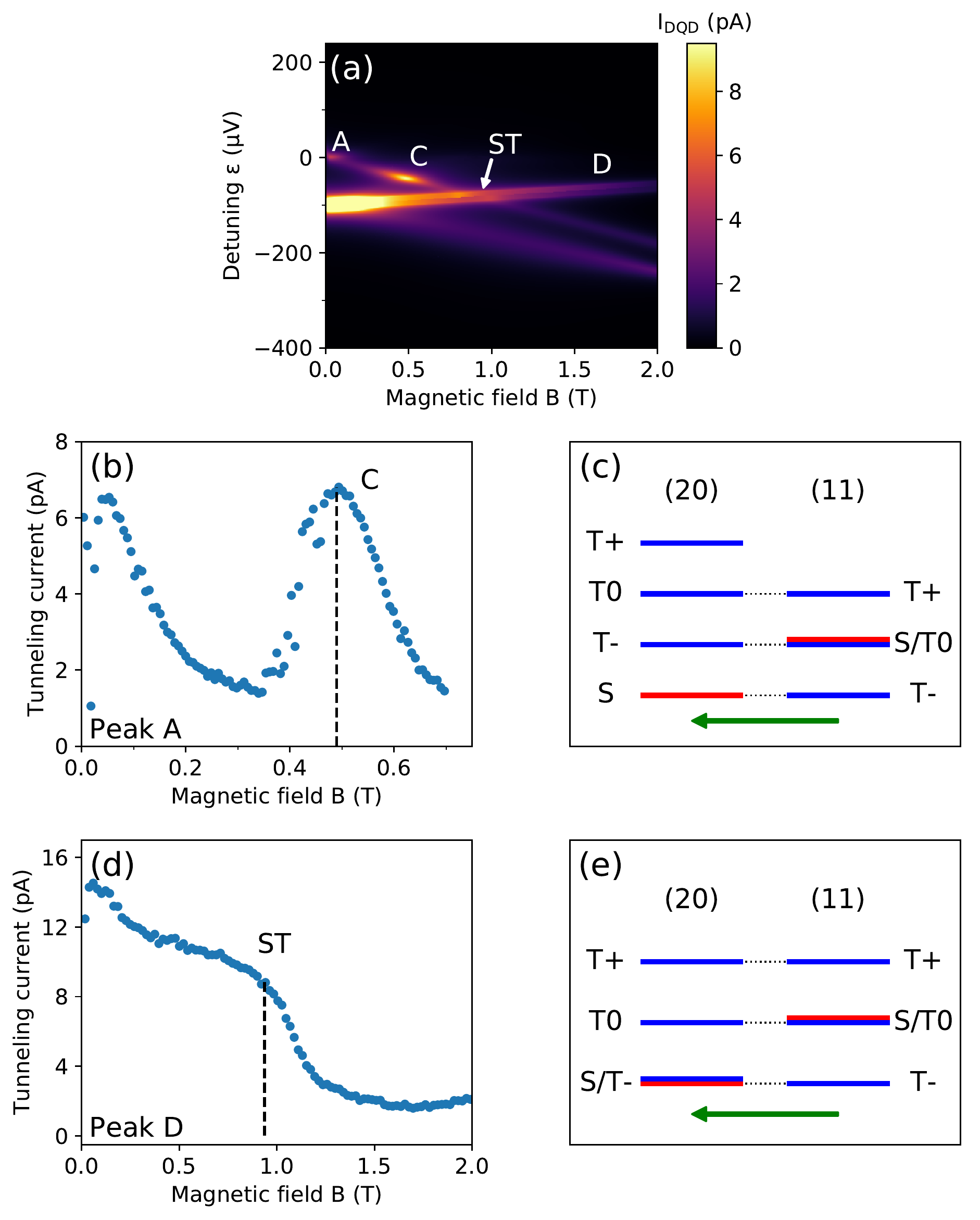}
    \caption{
    (a) Calculated tunneling current as a function of the detuning
    and the magnetic field, with fitted model parameters following the lines in Fig.~\ref{fig7}(e) and (f). 
    (b) Experimental amplitude of the current along the resonance A
    extracted from Fig.~\ref{fig7}(a) as a function of the magnetic field;
    the hotspot C occurs at $B\approx 0.5$ T.
    (c) Alignment of levels corresponding to the hotspot C.
    The green arrow shows the current direction.
    (d) The tunneling current on resonance D measured in the experiment 
    [Fig.~\ref{fig7}(a)];
    the point marked as ST corresponds to the singlet-triplet transition
    in the (20) charge configuration.
    (e) Alignment of levels corresponding to the singlet-triplet transition ST.
    The green arrow shows the current direction.
    }    
    \label{fig8}
\end{figure}

Next, we generate the theoretical plot reproducing the 
experimental dataset shown in Fig.~\ref{fig7}(a).
From the differential current in Fig.~\ref{fig7}(b) we find
the singlet-triplet transition in this experiment at the field 
$B_{ST} = 0.93$ T.
Moreover, the singlet-triplet gap appears to change somewhat nonlinearly
with the magnetic field, which we approximate by a linear
dependence of the $T_-(11)$--$S(20)$ resonance A, as shown in Fig.~\ref{fig7}(b)
with the green line.
Taking this dependence, we extract
$\Delta E_{ST}(B=0) = 99.5$ $\mu$eV and 
$\Delta E_{ST}(B=B_{ST}) = 80.8$ $\mu$eV and arrive at the relationship
$\Delta E_{ST}(B) = (99.5 - 20.1 B)$ $\mu$eV.
Also, we extrapolate the extracted dependencies of the model parameters
to the field values below $0.5$ T and assume the unfitted matrix element $t_N = t_N'$.
The simulation result is shown in Fig.~\ref{fig8}(a) on the same color scale
as the experimental data in fig.~\ref{fig7}(a).
We find that, besides the linearized magnetic-field dependence of the positions of the
resonant tunneling peaks, all features appearing in the experimental data
are well reproduced by our theoretical model.

\subsection{Signatures of level degeneracies (hot and cold spots)}

Now we focus on two current special points, marked in Fig.~\ref{fig8}(a)
as C and ST.
The hotspot C is found along the resonance A (see Fig.~\ref{fig5} for the
generic alignment of levels) and is clearly visible in Fig.~\ref{fig8}(b) 
showing the measured current extracted along that line.
This feature is detected at the magnetic field $B\approx 0.5$ T, at which
the alignment A coincides with the alignment C, as shown in Fig.~\ref{fig8}(c).
Here, the Zeeman energy is equal to exactly half
of the singlet-triplet gap $\Delta E_{ST}$,
which is measured between the configurations
$S(20)$ and $T_0(20)$.
At the magnetic fields away from that point, in either A or C there are always
levels of the (11) configuration which are not resonant with any of the
$(20)$ levels.
As a result, the tunneling current is limited by the spin relaxation process.
However, when the two alignments coincide, all $(11)$ levels are aligned with
their counterparts in the $(20)$ configuration, and therefore the current is only 
limited by the magnitude of tunneling elements, providing for a much more efficient
tunneling.
The second hotspot ST appears as a shoulder on the magnetic field dependence
of the current measured along the resonance D, as shown in Fig.~\ref{fig8}(d).
As is evident from the extracted field dependencies of the tunneling elements shown in
Fig.~\ref{fig7}(f), we would expect a smooth decay of the current
along that line, as it is governed by the spin-conserving matrix element $t_N'$.
However, in the vicinity of the singlet-triplet transition in the $(20)$ configuration
we detect an enhancement in the current.
This enhancement is brought about by another complete resonance of all $(11)$ levels,
as shown schematically in Fig.~\ref{fig8}(e).
Compared to the alignment of levels for the resonance D at an arbitrary field
[see Fig.~\ref{fig5}(d)], we see that at $B=B_{ST}$ the singlet $(20)$ level
(red) and the $(20)$ triplet $T_-$ (blue) together are in resonance with the 
$(11)$ triplet $T_-$.
This opens an additional conducting channel, which is absent at arbitrary fields
and leads to the observed enhancement of the current.
Both hotspots are reproduced in the model calculations.

\subsection{Equivalence of transport spectra for opposite tunneling direction}

\begin{figure}[t]
    \includegraphics[width=0.7\textwidth]{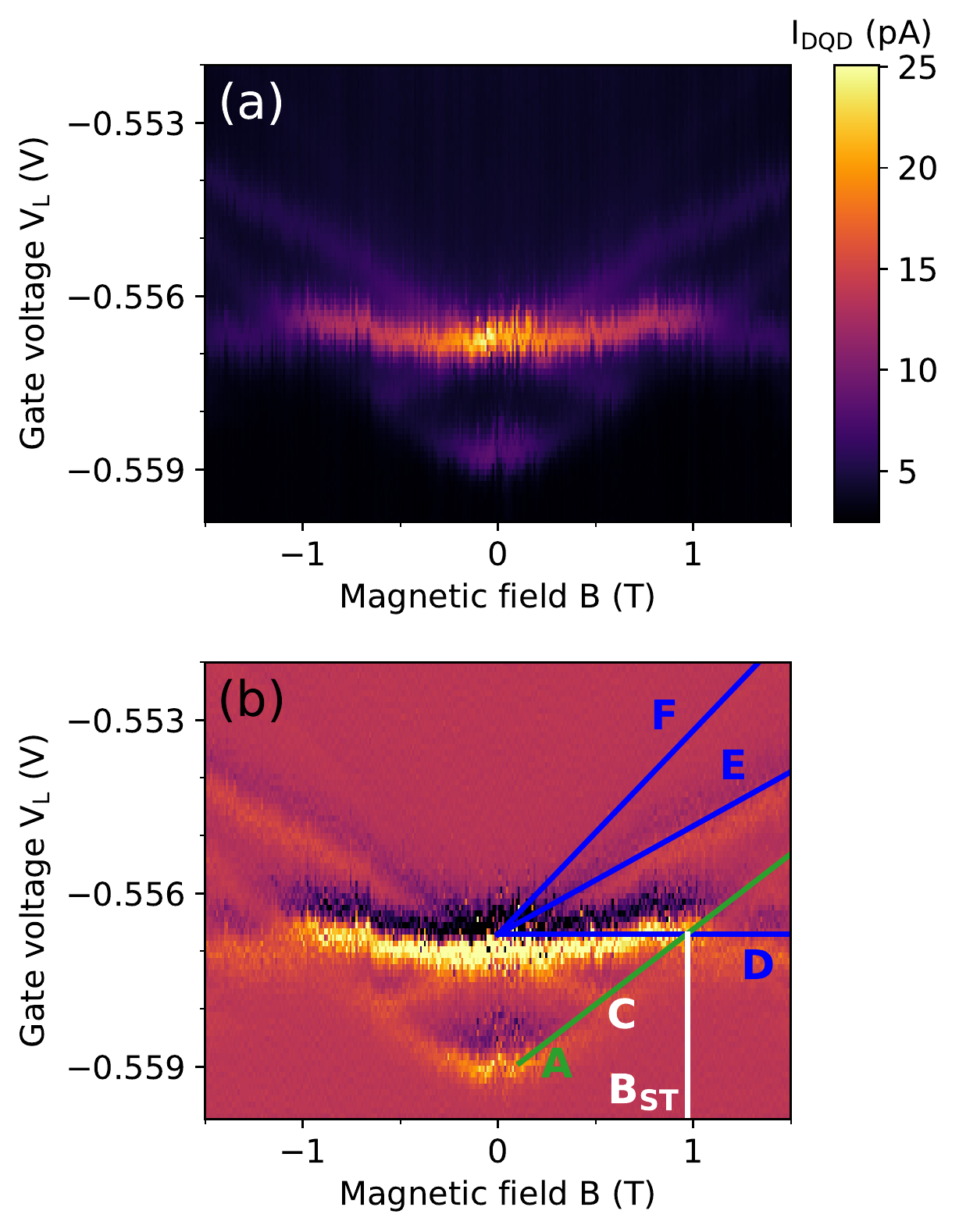}
    \caption{Current through the device (a) and the current derivative 
    $dI/dV_L$ (b)
    as a function of the magnetic field and the left gate voltage $V_L$
    measured in the blockaded direction 
    The interdot barrier is set up by applying the voltage $V_C = -0.09$ V.
    The transport spectrum is measured in the region of gate voltages
    within the green rectangle of Fig.~\ref{fig1}(c), i.e., in the direction
    opposite to that in Figs.~\ref{fig6} and~\ref{fig7}.
    }    
    \label{fig9}
\end{figure}

To conclude the discussion of transport in the blockaded direction,
in Fig.~\ref{fig9}(a) and (b) we show respectively the current and
the current derivative $dI/dV_L$ recorded at the region of voltages
within the green rectangle of Fig.~\ref{fig1}(c).
In this region, the transport occurs in the sequence
of charge configurations $(01)\rightarrow (11)\rightarrow (02) \rightarrow (01)$,
i.e., in the mirror opposite of the alignment shown in Fig.~\ref{fig2}(a).
The crucial difference here is that the "spectator" hole now resides in the
right-hand dot and the source-drain bias $V_{SD}$ is reversed.
Therefore, by adjusting the left gate voltage $V_L$ we are influencing the energies
of the states of the $(11)$ charge configuration more than we do those of 
the $(20)$ charge configurations.
As a result, in this alignment the spectra have to be recorded by
adjusting the voltage $V_L$ to the less negative values, as opposed to
Figs.~\ref{fig6} and~\ref{fig7}, and the characteristic transport peaks
appear inverted relative to the ones discussed earlier.
This is exactly what we observe in Fig.~\ref{fig9}.
With that in mind, we clearly identify the relevant resonance lines:
A, D, E, and F, with the singlet-triplet transition in the $(02)$ configuration
occurring at the magnetic field $B_{ST}\approx 1.0$ T.
We also recover a faint signature of the hotspot C.
We find that the transport spectra recorded in the blockaded
direction are qualitatively the same irrespective of the variant of the
tunneling sequence, i.e., they do not depend upon whether the spectator
hole resides in the left- or the right-hand dot.
Therefore, the tunneling peaks analyzed in this Section result
from the electronic properties of the two-hole double-dot
rather than from the details of hole-lead tunneling, and are
accurately and quantitatively reproduced by our theoretical model.

\section{Transport spectra of two holes in the
non-blockaded direction}

In this Section, we focus on the transport spectra in the "non-blockaded"
or conduction direction, i.e., the alignment of voltages in which
the sequence of charge configurations is $(10)\rightarrow(20)
\rightarrow(11)\rightarrow(10)$.
Just as it was in the discussion of the blockaded direction (Sec. IV A--C), the 
spectator hole is confined in the left dot.
However, now the first tunneling event brings an additional hole
from the left lead into the left dot, as depicted in Fig.~\ref{fig2}(b).
We note that we remain in the same region of the charging diagram,
marked in Fig.~\ref{fig1}(c) with the red rectangle, the only difference
being the direction of the source-drain voltage $V_{SD}$.
The theoretical model describing transport in the conduction direction
differs from the one described above only in one aspect, i.e.,
the coupling to leads.
Since the source and drain now correspond to the left and right
lead, respectively, we need to interchange the couplings
$\Gamma^{(in)}_i$ and $\Gamma^{(out)}_i$ in Equations (\ref{eq11})--(\ref{eq16}).
Owing to the formulation of our model in terms of molecular
orbitals, the entire density matrix formalism remains unchanged,
including the expression for the total current,Eq.~(\ref{tot_current}),
albeit with the new definition of the rate $\Gamma^{(out)}_i$.
The transport spectra recorded in the conduction direction
were already shown in Fig.~\ref{fig2}(d)
for the current, and Fig.~\ref{fig2}(f) for the current derivative.
We will now discuss them in detail.

\subsection{Positions of current maxima}

In contrast to the blockaded direction, here we use the levels
of the $(20)$ charge configuration as a spectroscopic tool
to map out the energy levels of the $(11)$ configuration.
To this end, we will now tune the left gate voltage $V_L$
towards more positive values.
This shifts the $(20)$ energies up relative to those of the
$(11)$ configuration.

We first perform a qualitative analysis assuming the fast spin-flip
relaxation, i.e., we will assume that only the lowest-energy
$(20)$ level is occupied.
As demonstrated in the previous Section, 
the doubly-occupied system undergoes the singlet-triplet transition
at the magnetic field $B=B_{ST}$.
Therefore, it is necessary to develop our analysis separately
for the magnetic fields smaller and larger than $B_{ST}$.

\begin{figure}[t]
    \includegraphics[width=0.7\textwidth]{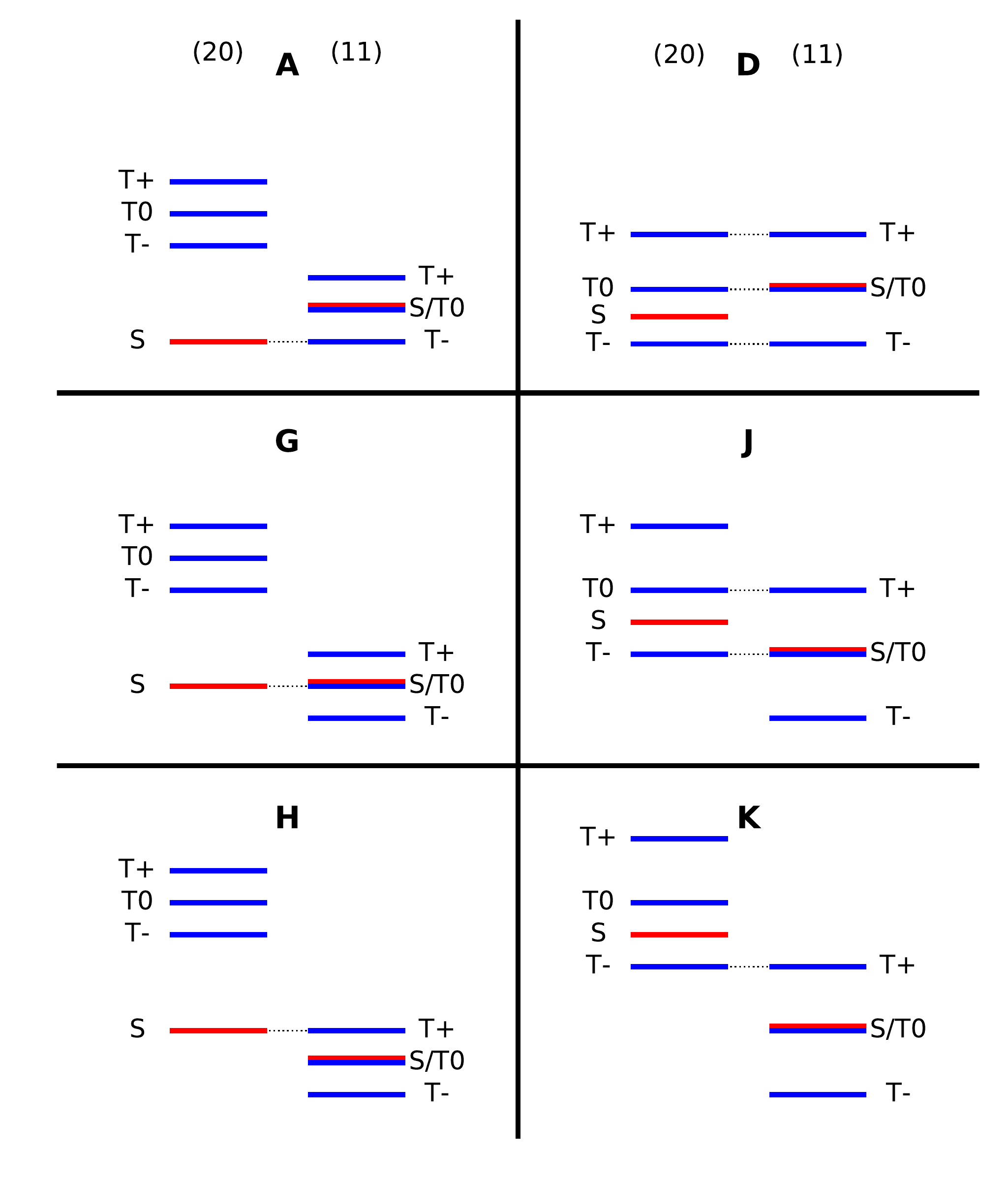}
    \caption{Different realizations of tunneling resonances between 
    $(20)$ (left in each panel) and $(11)$ (right) charge 
    configurations relevant for the conduction direction.  
    In the left-hand (right-hand) column we show
    the alignment for low, $B<B_{ST}$ (high, $B>B_{ST}$) magnetic field.
    Panels from top to bottom show level 
    alignments  with  increasing  positive
    detuning.   Red  (blue)  color  corresponds  to  singlet  
    (triplet) levels.    }    
    \label{fig10}
\end{figure}

Let us first focus on the region $B<B_{ST}$,
shown in the left-hand column of Fig.~\ref{fig10}.
Here the lowest-energy $(20)$ configuration is the singlet $S(20)$,
and its energy will now be compared to those of the $(11)$ configurations.
In Fig.~\ref{fig10}A we show the fundamental resonance, 
$S(20)$-$T_-(11)$, corresponding to precisely the same level alignment as
that described in Fig.~\ref{fig5}A for the blockaded direction.
As discussed before, the transport maximum for this resonance
will occur for the detuning $\Delta\varepsilon_A = -E_Z$.
As we shift the  $(20)$ energies higher, we encounter the second
resonance, $S(20)$-$S(11)$, shown in Fig.~\ref{fig10}G.
Note that only the resonance of the two singlets produces tunneling,
as the tunnel coupling between $S(20)$ and $T_0(11)$ is zero.
Since this resonance does not involve a spin flip, its position
in the spectra will be independent of the magnetic field, and will
correspond to the detuning $\Delta\varepsilon_G = 0$.
The third strong resonance will connect $S(20)$ with $T_+(11)$,
as shown in Fig.~\ref{fig10}H.
This spin-flip resonance will correspond to the detuning $\Delta\varepsilon_H = +E_Z$.
As we can see, at low magnetic fields we expect three main current
maxima, separated from one another by the Zeeman gap.

Let us now move on to the high magnetic fields, $B>B_{ST}$ (right-hand column of Fig.~\ref{fig10}).
Here, the lowest-energy $(20)$ configuration is the polarized
triplet $T_-(20)$.
The first, fundamental resonance will occur when this level
is aligned with the polarized triplet $T_-(11)$, as shown
in Fig.~\ref{fig10}D.
This alignment of levels has already appeared in the analysis
and is shown in Fig.~\ref{fig5}D, although for much lower magnetic 
fields.
We find this resonance at the detuning 
$\Delta\varepsilon_D = -\Delta E_{ST}(B)$.
Since the singlet-triplet gap $ \Delta E_{ST}(B)$ depends somewhat
on the magnetic field due to the diamagnetic effect, we expect that the 
position of this current
maximum will be weakly dependent on the field.
Moreover, since here we deal with alignment of all three triplets,
we expect that the amplitude of this resonance will be high.
As the detuning is made more positive, we find the next resonance
at the alignment shown in Fig.~\ref{fig10}J.
Here the energy matching occurs between $T_-(20)$ and $S(11)$
and between $T_0(20)$ and $T_+(11)$.
We therefore expect a current maximum at the detuning
$\Delta\varepsilon_J = -\Delta E_{ST}(B) + E_Z$.
As we align only two triplet levels, we expect that the amplitude
of this maximum will be smaller than that of the alignment D.
Finally, the third resonance occurs at the alignment of levels
$T_-(20)$ and $T_+(11)$, as shown in Fig.~\ref{fig10}K.
This alignment corresponds to the detuning 
$\Delta\varepsilon_K = -\Delta E_{ST}(B) + 2E_Z$.
We note that here we match the energies of only one pair of triplets,
and the transition itself involves a double spin flip, which
allows us to expect the current maximum with the lowest peak amplitude.
To summarize, in the region of high magnetic fields $B>B_{ST}$ we also expect
three current peaks separated from one another by the Zeeman gap.
Unlike in the low magnetic case, however, their positions
will also be influenced by the magnetic field-dependent 
singlet-triplet gap $\Delta E_{ST}$.

\begin{figure}[t]
    \includegraphics[width=0.7\textwidth]{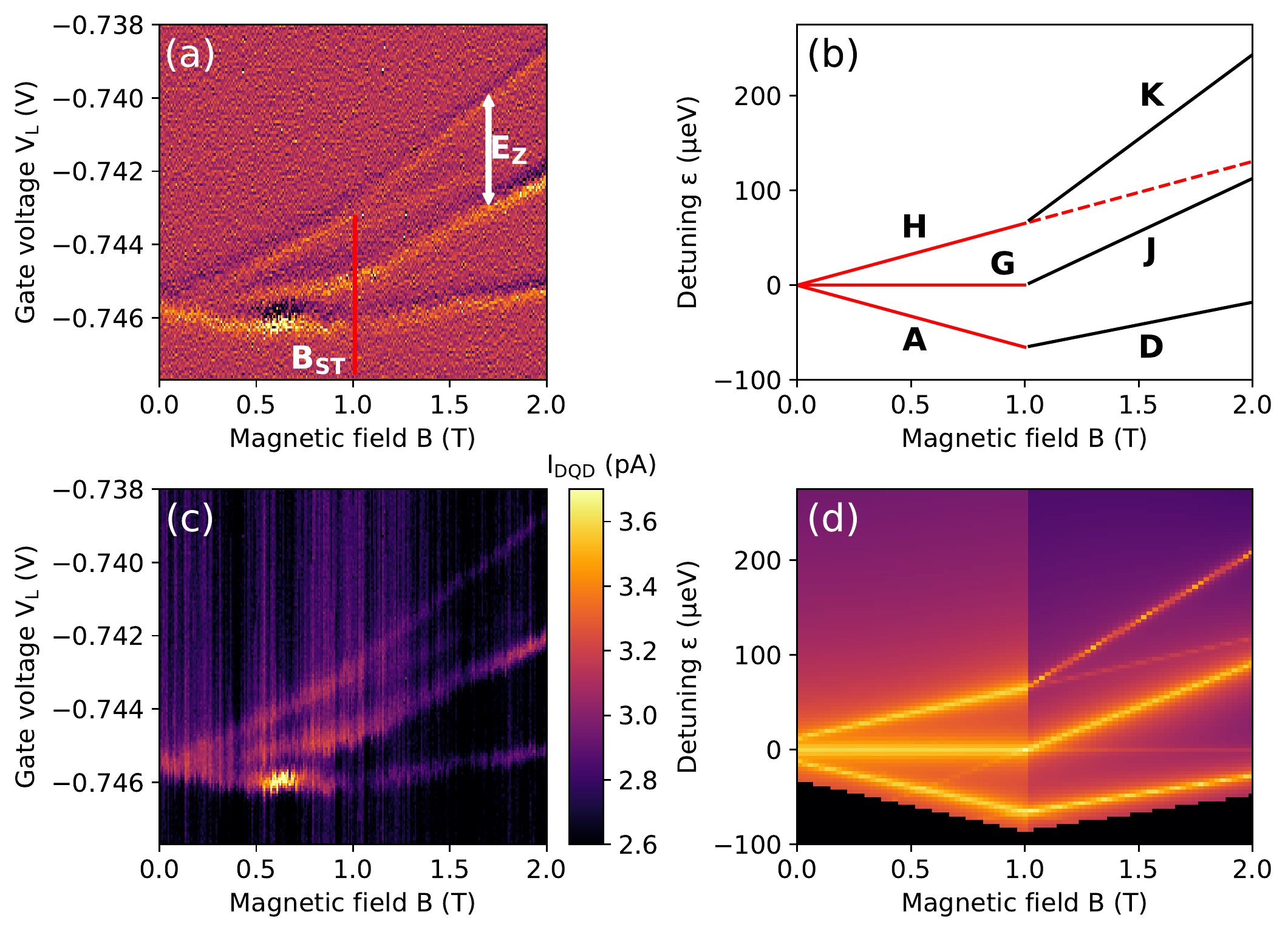}
    \caption{(a) Differential current $dI/dV_L$ as a function of the
    gate voltage $V_L$ and the magnetic field in the conduction direction
    for the tunnel barrier defined by the voltage $V_C=-0.07$ $V$.
    Note the identification of characteristic elements
    of the spectrum, i.e., the magnetic field
    $B_{ST}$ corresponding to the singlet-triplet transition 
    and the Zeeman energy $E_Z$.
    (b) Calculated positions of the transport maxima as a function of
    detuning     and the magnetic field. 
    Labels correspond to the resonances shown in Fig.~\ref{fig10}.
    Panels (c) and (d) show respectively the tunneling current 
    measured as a function of the gate voltage $V_L$ and the 
    magnetic field and the corresponding theoretical simulation.}
    \label{fig11}
\end{figure}

Figure~\ref{fig11}(a) shows the current derivative $dI/dV_L$ 
as a function of the gate voltage $V_L$ and the magnetic field 
in the conduction direction for the tunnel barrier defined by the 
voltage $V_C=-0.07$ $V$.
In Fig.~\ref{fig11}(b) we reproduce the positions of transport peaks
with our model and identify each resonance with the label from
Fig.~\ref{fig10}.
We find that the transition from the low-field to the high-field region,
occurring at the critical field $B_{ST}\approx 1$ T,
is marked by characteristic kinks in the resonances.
By tracking any gap between two adjacent maxima we can establish
the effective $g$-factor.
Indeed, at $B=B_{ST}$ we read the gaps to be $E_Z(B_{ST}) = 65.3$ $\mu$eV,
giving $g^*=1.13$.
Note that the value of $g^*$ is different than that found in the
opposite tunneling direction.
This is in line with the dependence of $g^*$ on gate voltages
reported by us recently.~\cite{Bogan2019-2}
Moreover, at the critical magnetic field the Zeeman gap is equal to
the singlet-triplet gap $\Delta E_{ST}$.
By tracing the slope of the line D we recover directly the 
magnetic field dependence of the singlet-triplet gap, which in the linear
approximation is $\Delta E_{ST}(B) = (112.9 -47.5 B)$ $\mu$eV.
These two parameters suffice to reproduce the entire transport spectrum.
We note that the low-field resonance H continues into the high-field
regime, even though the energy matching takes place between
excited states (singlet $S(20)$ and triplet $T_+(11)$) of 
both charge configurations.
This indicates that in the $(20)$ charge configuration both the
ground and excited states have non-zero occupation probabilities and 
can create resonances detected as transport peaks.

In a more precise analysis, we reproduce the measurement of
the tunneling current, shown in Fig.~\ref{fig11}(c),
with the full density-matrix simulation, whose results are
shown in Fig.~\ref{fig11}(d).
Here we use model (i.e., unfitted) tunneling matrix elements
identical to those given in Sec.~\ref{idealized}. 
The only exception is made for the spin-flip relaxation time,
which in this dataset is taken to be $T_{SF}=100$ ns for 
all magnetic fields.
The most striking feature in the theoretical plot is the
difference in the overall current amplitudes between the low-field
and high-field regions, seen as the difference in the background 
shade between $B<1$ T and $B>1$ T.
This is a direct consequence of the change of the ground-state
configuration in the $(20)$ charge state.
Indeed, for low fields the relaxation occurs from all triplets
into the singlet $S(20)$.
Since in our model the relaxation can happen only via
a spin-flip, such process will not connect the unpolarized 
triplet $T_0(20)$ to $S(20)$, as here the spin projection does not change.
The relaxation can still take place in an indirect fashion, i.e.,
by $T_0(20)$ relaxing to $T_-(20)$ and that to $S(20)$.
Such a slower relaxation channel becomes important with a 
sizeable starting occupation of $T_0(20)$.
On the other hand, in the high-field region the lowest-energy
configuration is $T_-(20)$ and therefore both spin-unpolarized
configurations, $S(20)$ and $T_0(20)$, are directly connected
to it via the spin-flip relaxation channel.
The double-flip relaxation is required from the highest-energy
configuration $T_+(20)$ to $T_-(20)$, but this indirect process
can be realized by two pathways, i.e., via either spin-unpolarized
configuration, and therefore is also fast.

Secondly, apart from all main maxima, we still track the secondary
features: the resonances G and H continuing to the high-field regime,
and the resonance J visible already in the low-field regime.
Not all these features are apparent in the experimental data.
In the next Section we will present a more detailed 
fitting of the tunneling matrix elements, resulting in a closer correspondence of
experimental and theoretical spectra.

\subsection{Tunneling amplitudes}

\begin{figure}[t]
    \includegraphics[width=0.7\textwidth]{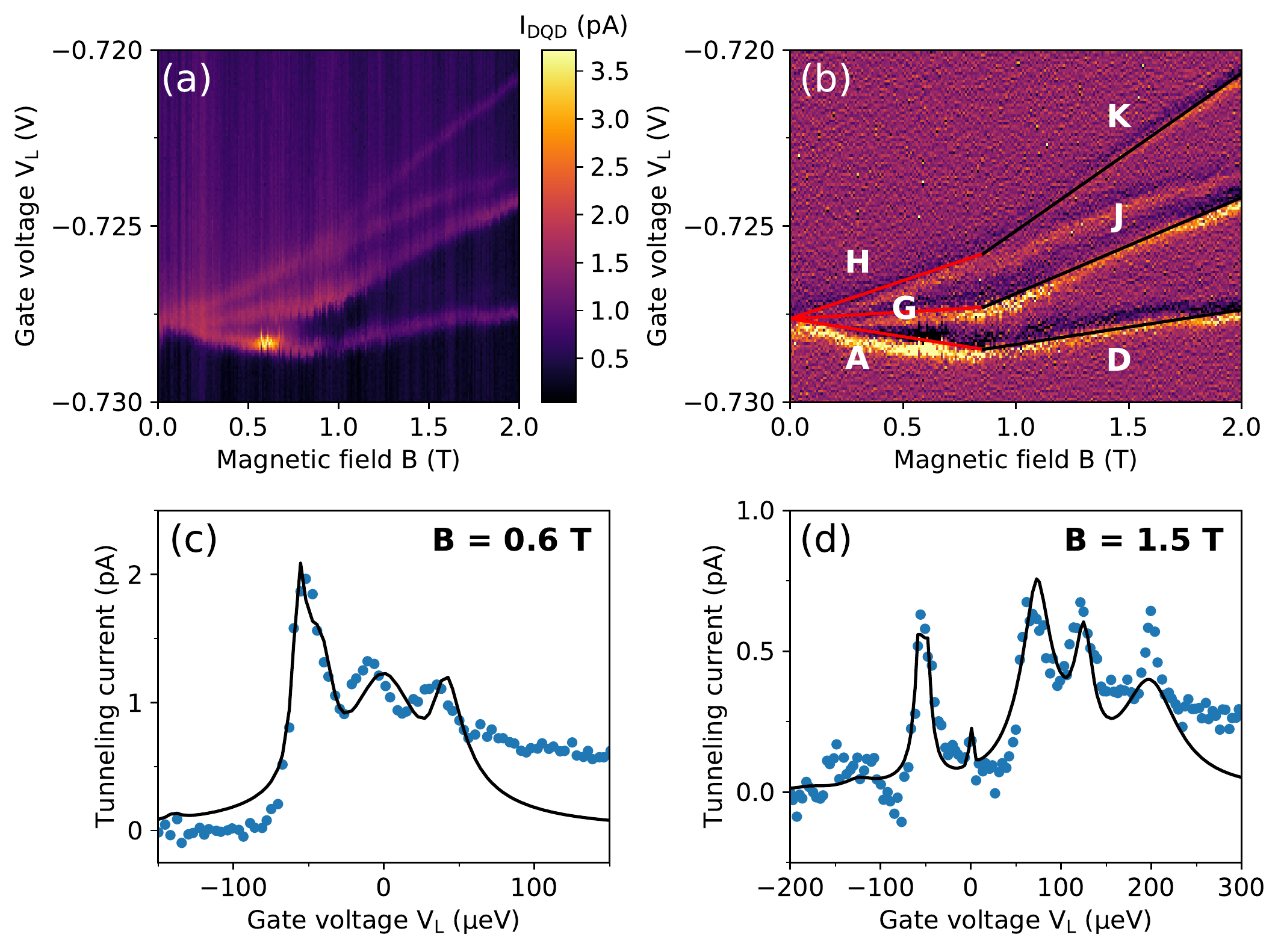}
    \caption{(a) Tunneling current and (b) the derivative $dI/dV_L$ 
    as a function of the gate voltage
    $V_L$ and the magnetic field with the interdot gate voltage
    $V_C=-0.09$ $V$. 
    Panels (c) and (d): the tunneling current as a function of the
    left gate voltage $V_L$ extracted from (a) at the magnetic field $B=0.6$ T
    and $1.5$ T, respectively; experimental data (symbols) and theoretical fits (lines).
    }    
    \label{fig12}
\end{figure}

For the quantitative analysis of the current spectra we take
the transport measurement performed at the interdot gate voltage
$V_C=-0.09$ $V$.
Figure~\ref{fig12}(a) shows the measured current, and Fig.~\ref{fig12}(b) 
-- the differential current $dI/dV_L$ as a function of the voltage $V_L$
and the magnetic field.
We perform the genetic fitting of all tunneling matrix elements,
including the double-spin flip element $\tau$, assuming the coupling
to the leads to be $\Gamma = 2$ GHz.
Figure~\ref{fig12}(c) and (d) shows the experimental data
(symbols) and the theoretical fit (lines) respectively for the
magnetic field $B=0.6$ T and $B=1.5$ T.
In contrast to the data in the blockaded direction, shown 
in Fig.~\ref{fig7}, here we deal with a higher interdot barrier,
resulting in a smaller current overall.
In consequence, the charge leakage visible at large gate voltages 
$V_L$ contributes more to the spectra.
Nonetheless, the positions and widths of transport peaks could be fitted
satisfactorily.
The maximum D in the high-field regime is treated in the same way as
in the blockaded direction, i.e., we truncate it manually 
to account for the decoherence in our system.

\begin{figure}[t]
    \includegraphics[width=0.7\textwidth]{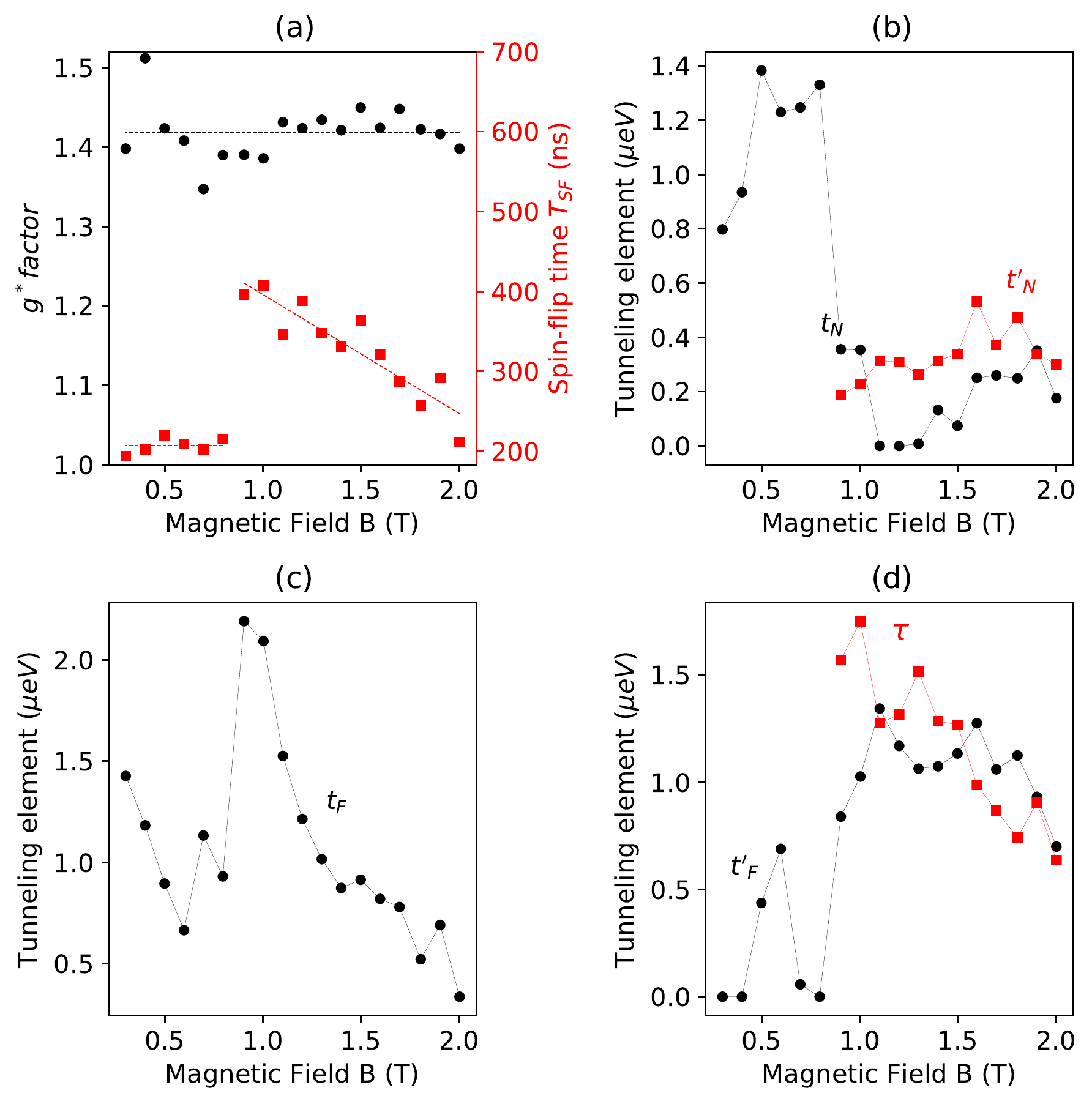}
    \caption{Results of fitting of model parameters in the conduction 
    direction as a function of the magnetic field. 
    (a) The effective $g^*$ factor (black) and the spin-flip relaxation
    time $T_{SF}$ (red).
    (b) The spin-conserving matrix elements $t_N$ (black) and $t_N'$   
    (red).
    (c) The spin-flip matrix element $t_F$.
    (d) The secondary spin-flip matrix element $t_F'$ (black) and
    the double-spin flip element $\tau$ (red).
    In panels (b)--(d), the lines are guides to the eye.
    }    
    \label{fig13}
\end{figure}

In Fig.~\ref{fig13} we show the model parameters extracted
in the fitting procedure as a function of the magnetic field.
As shown in Fig.~\ref{fig13}(a), the extracted $g^*$-factor 
is magnetic field-independent, with the average value of $g^*=1.42$.
The spin-flip time $T_{SF}$, on the other hand, is field-independent
only in the low-field section, with an approximate value of
$T_{SF}=207.1$ ns.
For the magnetic fields $B>B_{ST}$ the spin-flip time decreases
linearly with the field, with the fitted functional relationship
of $T_{SF}(B)= (545.0 - 148.9 B)$ ns.
We note that at the singlet-triplet transition, occurring at the 
magnetic field $B_{ST}=0.95$ T, the relaxation time appears to jump
discontinuously.
In our model this discontinuity appears as
a result of the change of the ground state in the $(20)$ configuration and the concomitant rearrangement of relaxation pathways, as discussed in the previous Section.
In fitting we also recover the magnetic-field dependence of the
singlet-triplet gap in the $(20)$ charge configuration (not shown).
This gap changes approximately linearly according to
$\Delta E_{ST}(B) = (126.9 - 44.7 B)$ $\mu$eV.

\begin{figure}[t]
    \includegraphics[width=0.7\textwidth]{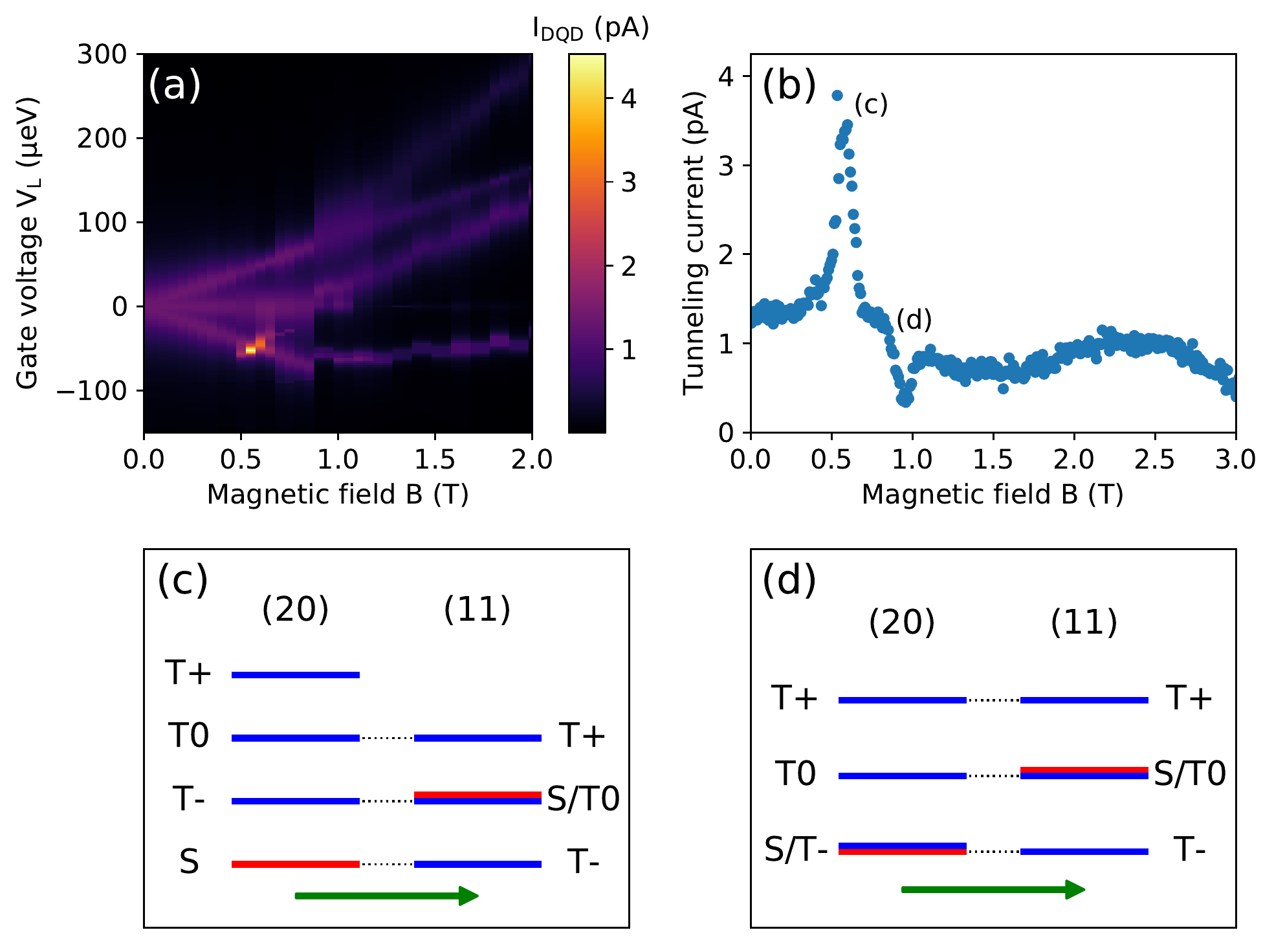}
    \caption{
    (a) Calculated tunneling current as a function of the detuning
    and the magnetic field, with fitted model parameters as in
    Fig.~\ref{fig13}.
    (b) The tunneling current extracted from the experimental data as in
    Fig.~\ref{fig12}(a) along the lowest maximum (A at low fields
    and D at higher fields).
    (c) Alignment of levels corresponding to the maximum in tunneling
    current in panel (b).
    The green arrow shows the current direction.
    (d) Alignment of levels corresponding to the narrow minimum
    in panel (b).     The green arrow shows the current direction.
    }    
    \label{fig14}
\end{figure}

Figure ~\ref{fig13} also shows the tunneling matrix elements extracted 
in fitting. 
In Fig.~\ref{fig13}(b) we show the extracted magnetic field dependence 
of the spin-conserving elements $t_N$ (black) and $t_N'$ (red);
Fig.~\ref{fig13}(c) shows the spin-flip element $t_F$,
and Fig.~\ref{fig13}(d) shows the spin-flip element $t_F'$ 
and the double-spin-flip element $\tau$.
In contrast to the results obtained in the blockaded direction
(Fig.~\ref{fig7}), here the elements do not present
simple exponential dependencies on the field.
In general, all elements do appear to diminish
at higher magnetic fields.
However, the elements $t_N$ and $t_F$ appear to change
discontinuously across the singlet-triplet transition at $B_{ST}$.
This discontinuity is again traced to the change of relaxation pathways, perturbing the timing of the tunneling processes.
Moreover, the large scatter in fitted elements is most likely
due to the level of noise evident in the experimental data of
Fig.~\ref{fig12}(a).
This is why we choose not to fit these magnetic-field dependencies
with any definite model function, but use this data point-by-point
in our calculations.
The result of our model calculations based on the extracted parameters is presented in Fig.~\ref{fig14}(a).
As we can see, the calculation with fitted parameters faithfully
reproduces the experimental spectra.

\subsection{Signatures of level degeneracies (hot and cold spots)}

Let us now analyze the current maximum corresponding to
the most negative detuning, i.e., along the resonance A
in low fields, and along the resonance D in high fields.
The current amplitudes corresponding to these resonances, extracted
from the experimental data of Fig.~\ref{fig12}(a), are
shown in Fig.~\ref{fig14}(b).
We note that we have already made a similar analysis for the
blockaded direction, with the corresponding results shown in
Fig.~\ref{fig8}.
Here we focus on two features: the peak at the magnetic field
$B\approx 0.53$ T and the sharp minimum at $B=0.95$ T.

The alignment of levels corresponding to the current peak is
shown in Fig.~\ref{fig14}(c).
This alignment appears at the magnetic field for which
the Zeeman energy $E_Z$ is equal to half of the singlet-triplet
gap $\Delta E_{ST}$ in the $(20)$ charge configuration.
At this field, three levels: $S(20)$, $T_-(20)$, and $T_0(20)$
are aligned respectively with $T_-(11)$, $S(11)$ degenerate with 
$T_0(11)$, and $T_+(11)$.
The only $(20)$ level not on resonance with any of the $(11)$ levels
is $T_+(20)$.
In such an alignment we deal with three conduction channels,
and, with the exception of $T_+(20)$, we do not expect any
charge accumulation to occur.
The highest $(20)$ level is efficiently emptied to lower levels
by the spin-flip relaxation along two channels: to $S(20)$ and
$T_0(20)$.
This is in contrast to alignments at different fields, shown generically
in Fig.~\ref{fig10}, where only one $(20)$ level is in resonance with
one of the $(11)$ levels, resulting in a more substantial charge
accumulation on other $(20)$ levels and more complex relaxation
pathways.
As a result, we observe a strong peak in the current, 
similarly to the maximum discussed in Fig.~\ref{fig8}(b) and (c).

The current minimum, seen in Fig.~\ref{fig14}(b) at
$B_{ST}=0.95$ T, on the other hand, does not match any feature
in the blockaded direction.
On the contrary, in Fig.~\ref{fig8}(d) at the singlet-triplet transition
we observe a pronounced high-current shoulder.
However, in both the blockaded and conduction directions
the level alignment is the same, as shown in Fig.~\ref{fig8}(e)
and Fig.~\ref{fig14}(d), respectively.
To understand this difference, we write a reduced Hamiltonian
at $B=B_{ST}$, involving only three resonant levels: $S(20)$,
$T_-(20)$, and $T_-(11)$, in that order.
Extracting these three basis states from the general Hamiltonian,
Eq. (\ref{th_hamil}), we have:
\begin{equation}
    \hat{H}_{2H}(B=B_{ST}) = \left[
    \begin{array}{ccc}
    E_0 & 0 & it_F \\
    0 & E_0 & -t_N' \\
    -it_F & -t_N' & E_0 
    \end{array}
    \right].
    \label{ham_reduced}
\end{equation}
Since the three levels are on resonance, at this magnetic field
$E_0 = E_{S(20)} = E_{T(20)} - E_Z = E_{T(11)} - E_Z$.
One of the eigenstates of this Hamiltonian has the form
\begin{equation}
|B\rangle = \frac{1}{\sqrt{t_F^2 + (t_N')^2}}
\left( t_N' |S(20)\rangle - it_F |T_-(20)\rangle\right),
\end{equation}
with the eigenenergy $E_0$.
This state is composed only of the levels of the $(20)$ charge
configuration.
We stress that both the spin conserving and spin flip tunneling 
processes contribute to its formation.
The two other eigenstates are superpositions of levels
both of $(20)$ and $(11)$ type.
Let us now analyze the tunnel connection of the state $|B\rangle$
to the leads.
In the blockaded direction, $|B\rangle$ is not connected directly 
to the source, since it does not contain admixtures of the $(11)$
charge configurations.
The occupation of this level may occur only by spin-flip relaxation,
whereupon $|B\rangle$ is efficiently emptied into the drain owing to
its good connection with that lead via the $(20)$ levels.
The two remaining eigenstates are connected both to the source and the
drain, and therefore form efficient transport channels.
As a result, no charge accumulation occurs on any levels shown
in Fig.~\ref{fig8}(e), and this multiple resonance is seen as a 
maximum in the current seen in Fig.~\ref{fig8}(d).

On the other hand, in the conduction direction the state $|B\rangle$
is filled from the source owing to contributions of the $(20)$
configuration, but cannot empty directly into the drain,
as it lacks admixtures of the $(11)$ charge configuration.
As a result, $|B\rangle$ is now a blocking eigenstate.
Even though at $B=B_{ST}$ all available levels are in resonance,
the overall current will be suppressed and the measured amplitude
will be determined by the spin-flip relaxation time.
We note that the state $|B\rangle$ is an exact eigenstate only at 
$B=B_{ST}$.
As we depart from that magnetic field, the eigenstate of
the Hamiltonian (\ref{ham_reduced}) corresponding to $|B\rangle$
will acquire an increasing contribution of the $(11)$ charge
configuration, and therefore will become an increasingly
efficient transport channel, which is seen as an increase of the current
both below and above $B_{ST}$.

\subsection{Equivalence of transport spectra for opposite tunneling direction}

\begin{figure}[t]
    \includegraphics[width=0.7\textwidth]{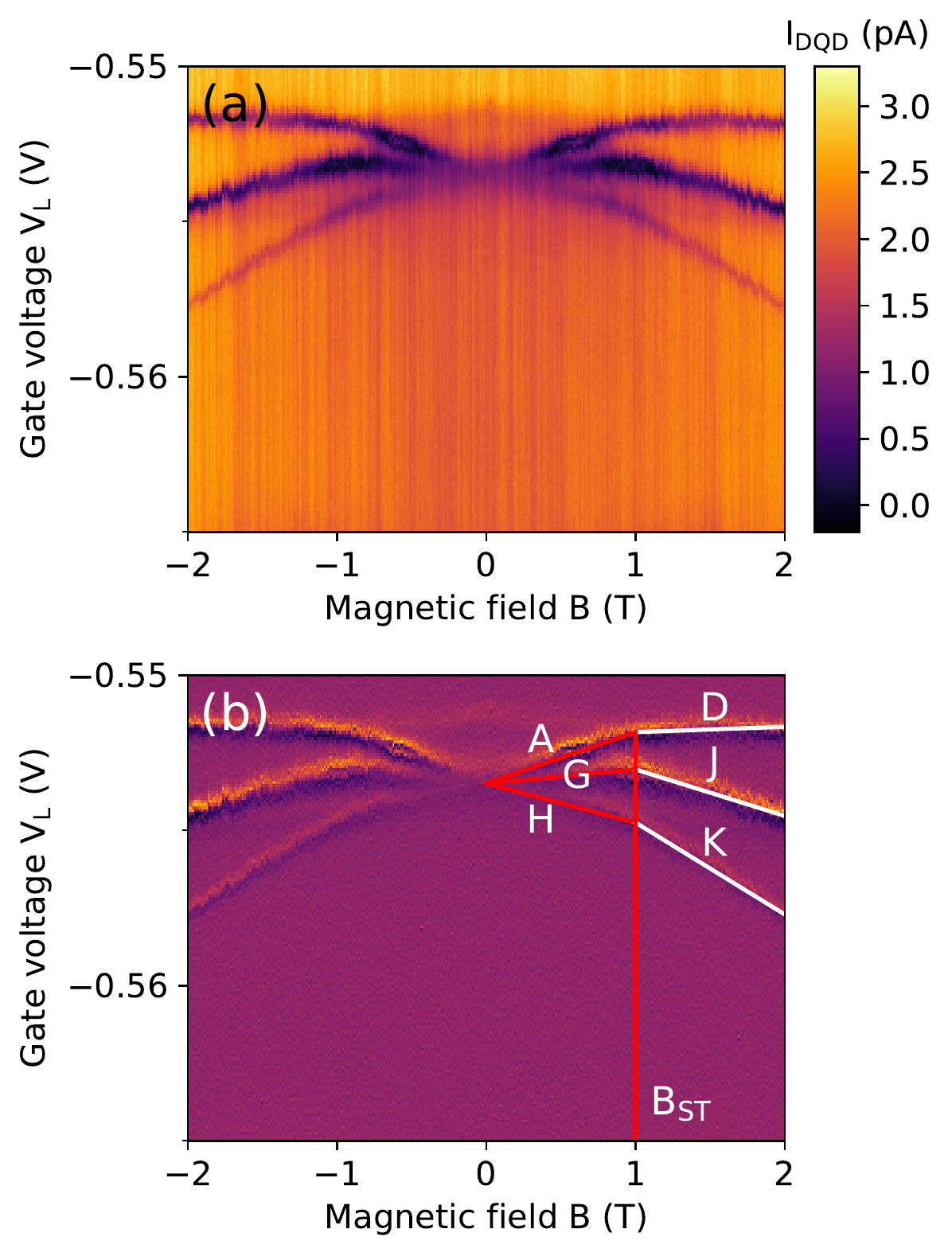}
    \caption{Current (a) and the current derivative $dI/dV_L$ (b)
    as a function of the magnetic field and the left gate voltage $V_L$
    measured in the conduction direction 
    The interdot barrier height is defined by the voltage $V_C = -0.09$ V.
    The transport spectrum is measured in the region of gate voltages
    within the green rectangle of Fig.~\ref{fig1}(c), i.e., in the direction
    opposite to that in Figs.~\ref{fig11} and~\ref{fig12}.
    }    
    \label{fig15}
\end{figure}

To confirm our analysis, in Fig.~\ref{fig15}(a) and (b) we show 
respectively the current and
the current derivative $dI/dV_L$ recorded in the conducting direction
at the region of voltages within the green rectangle of
Fig.~\ref{fig1}(c).
As discussed already for the blockaded direction,
here the spectator hole resides in the right-hand dot and the
source-drain voltage $V_{SD}$ is reversed relative to the
alignment discussed above.
Since now changing the voltage $V_L$ influences the $(11)$
charge configurations more than it does the $(02)$ configurations,
the series of resonances occurs as we tune $V_L$ towards
the more negative values, and the pattern of transport peaks is
reversed compared with Fig.~\ref{fig12}(a).
Informed by that, we identify all principal transport features
and label them as in Fig.~\ref{fig15}(b).
The characteristic kink in the spectra, corresponding to the
singlet-triplet transition in the $(20)$ configuration,
is seen at $B\approx 0.9$ T.

\section{Conclusions}

In conclusion, we have reported and analyzed experimentally and theoretically
the magneto-transport spectra of two holes in a gated lateral GaAs double dot
system.
The system was probed with transport spectroscopy in the high 
source-drain voltage regime.
Owing to the strong spin-orbit interaction, the measured current
revealed a series of maxima corresponding to both spin-conserving
and spin-flip resonances, and therefore allowed to map out
the complete energy spectra of the system of two confined holes.
Depending on the tunneling direction, defined by the sign of the
source-drain voltage, we analyzed the singlet and triplet states
of the doubly-occupied $(20)$ charge configuration (the blockaded
direction) as well as the singly-occupied $(11)$ configuration
(the conduction direction) as a function of the magnetic field.

In the blockaded direction we did not find the hole current
suppression accompanying the Pauli spin blockade observed in gated devices
confining electrons.
Instead, we found enhancements of tunneling current 
at the gate voltages and magnetic field for which multiple
levels were aligned, i.e., on energy resonance.
In the conduction direction one of such alignments resulted in
a strong suppression of the current due to the formation of a
strongly blockaded excited state in the spectra.
This state was found to be a consequence of an interplay of
the spin conserving and spin-flip tunneling.

We have formulated a detailed theoretical model, accounting for
both spin-conserving and spin-flip tunneling processes as well as the 
tunnel connection to the leads
and the spin-flip relaxation process.
We demonstrated that this model allows for quantitative
fitting to experimental tunneling current spectra as a function of
interdot detuning and the magnetic field.
The fitting allowed to extract the dependencies of the tunneling matrix 
elements on the magnetic field.

\section*{Acknowledgment}
A. B. and S. S. thank NSERC for financial support.
This work was performed, in part, at the Center for Integrated Nanotechnologies, a U.S. DOE, Office of Basic Energy Sciences, user facility. Sandia National Laboratories is a multimission laboratory managed and operated by National Technology and Engineering Solutions of Sandia, LLC., a wholly owned subsidiary of Honeywell International, Inc., for the U.S. Department of Energy's National Nuclear Security Administration under contract DE-NA-0003525. The view expressed in the article do not necessarily represent the views of the U.S. Department of Energy or the United States Government.

\appendix
\section{Tunnel coupling with the leads \label{current_conditions}}
\label{app:leads}

The tunneling of the hole across our system consists of two phases: 
(i) tunneling of a single hole from the source (the right-hand lead) into the double-dot 
already containing one "spectator" hole, and (ii) tunneling of the single hole into the 
drain (the left-hand lead) leaving the "spectator" hole behind.
In all these processes we assume that the "spectator" hole occupies one of the levels
$|L1\Downarrow\rangle$ or $|L1\Uparrow\rangle$.
In principle, the tunneling of the second hole may leave behind
the "spectator" hole in the state $|L2\Downarrow\rangle$ or
$|L2\Uparrow\rangle$.
However, this would necessitate a transfer of energy of order of
$\varepsilon_{L2}-\varepsilon_{L1}$ from the leads. 
We assume that our choice of $V_{SD}$ does not allow this.

The quantitative description is cast in the language of elementary tunneling rates,
which, in terms of non-hybridized single-hole levels, are denoted as
$\Gamma^{(in)}$ and $\Gamma^{(out)}$.
The former parameter quantifies the tunneling from the source lead
into the right dot, depicted in Fig.~\ref{fig3}(d) and (e) by the curved arrows, 
respectively red for the spin up, and blue for the spin down.
The latter parameter is visualized in Fig.~\ref{fig3}(b) and (c) by the corresponding
arrows showing the tunneling from the left dot into the drain.
We assume that the drain is coupled equally to the states $|L1\sigma\rangle$ and
$|L2\sigma\rangle$.

We now calculate the tunneling rates involving the hybridized two-hole states
$|i\rangle$ (\ref{hybrid-2h}) using the procedure outlined in Ref.~\onlinecite{Qassemi2009}.
We define the hole spin-, state-, and parameter-dependent tunneling 
rates 
\begin{eqnarray}
\Gamma_i^{(in)}(\sigma_1,\sigma_2) &=& \Gamma^{(in)}\sum_{j=1}^8 |A_j^{(i)}|^2 
|\langle i|h^+_{R\sigma_1} h^+_{L1\sigma_2}|0\rangle|^2,\\
\Gamma_i^{(out)}(\sigma_1,\sigma_2) &=& \Gamma^{(out)}\sum_{j=1}^8 \sum_{K=L1}^{L2} |A_j^{(i)}|^2 
|\langle 0|h_{L1\sigma_2} h_{K\sigma_1}|i\rangle|^2,
\end{eqnarray}
where $\sigma_1$ denotes the spin of the hole which tunnels into or out of the system, 
while $\sigma_2$ denotes the spin of the resident "spectator" hole.
For the two-hole state $|i\rangle$ to contribute to the total current,
both of these rates have to be nonzero, although not necessarily for the
same choice of $\sigma_1$ and $\sigma_2$, i.e., we will consider both the
spin-conserving and spin-flip conduction channels.
Considering all combinations of the two spin indices, we arrive at the
following tunneling rates:
\begin{eqnarray}
\Gamma_i^{(in)}(\Downarrow,\Downarrow) &=& \Gamma^{(in)}|A^{(i)}_5|^2,
\label{eq11}\\
\Gamma_i^{(in)}(\Uparrow,\Downarrow) &=& \Gamma_i^{(in)}(\Downarrow,\Uparrow) =
\frac{\Gamma^{(in)}}{2}\left(|A^{(i)}_6|^2+|A^{(i)}_7|^2\right),\\
\Gamma_i^{(in)}(\Uparrow,\Uparrow) &=& \Gamma^{(in)}|A^{(i)}_8|^2,\\
\Gamma_i^{(out)}(\Downarrow,\Downarrow) &=& \Gamma^{(out)}|A^{(i)}_2|^2,\\
\Gamma_i^{(out)}(\Uparrow,\Downarrow) &=& \Gamma_i^{(out)}(\Downarrow,\Uparrow) =
\frac{\Gamma^{(out)}}{2}\left(2|A^{(i)}_1|^2+|A^{(i)}_3|^2\right),\\
\Gamma_i^{(out)}(\Uparrow,\Uparrow) &=& \Gamma^{(out)}|A^{(i)}_4|^2.
\label{eq16}
\end{eqnarray}
The asymmetry of the $\Gamma_i^{(in)}(\Uparrow,\Downarrow)$ and
$\Gamma_i^{(out)}(\Uparrow,\Downarrow)$ is understood from Fig.~\ref{fig3}: 
there is only one way to add a hole with the spin opposite to that
of the resident particle [one arrow in panel (d)], but there are two ways of removing
one hole from the $|S(20)\rangle$ configuration [two arrows in panel (b)].

\section{Spin-flip relaxation}
\label{app:relaxation}

For the single-dot system, the spin-flip relaxation is described in terms of the
spin relaxation time $T_1$ measured by us previously.~\cite{Bogan2019}
For example, with the "spectator" hole, any charge density corresponding 
to the occupation of the $|L\Uparrow\rangle$ state will decay to $|L\Downarrow\rangle$ 
at the rate $\Gamma_{SF} = {1}/{T_1}$.
In the more general treatment involving two holes, we assume that the 
spin-flip process does not lead to the redistribution
of charge, i.e., will not connect the $(20)$ configurations with the $(11)$ ones.
Second, we account for the phonon-mediated Dresselhaus spin-orbit interaction 
being the physical mechanism of the spin relaxation.~\cite{Bogan2019}
This mechanism connects configurations differing in their spin projections by one,
i.e., the two polarized triplets, $T_+$ and $T_-$, are both connected to the singlet
and unpolarized triplet $T_0$, but no other relaxation pathways are accounted for.
For simplicity, all pathways possible in our system are characterized by the same 
relaxation rate $\Gamma_{SF}$.

Rotation to the two-hole molecular basis, also accounting for the incoherent
character of the relaxation process, leads to the following generalized
relaxation rates between the two-hole molecular states $|i\rangle$ and $|j\rangle$:
\begin{eqnarray}
    \Gamma^{(SF)}_{ij} &=& \frac{1}{T_1}\left[
    \left( |A^{(i)}_1|^2 + |A^{(i)}_3|^2 \right)\left( |A^{(j)}_2|^2 + |A^{(j)}_4|^2 \right)
+   \left( |A^{(j)}_1|^2 + |A^{(j)}_3|^2 \right)\left( |A^{(i)}_2|^2 + |A^{(i)}_4|^2 \right)
\right.\nonumber\\
&+& \left.  \left( |A^{(i)}_6|^2 + |A^{(i)}_7|^2 \right)\left( |A^{(j)}_5|^2 + |A^{(j)}_8|^2 \right)
+   \left( |A^{(j)}_6|^2 + |A^{(j)}_7|^2 \right)\left( |A^{(i)}_5|^2 + |A^{(i)}_8|^2 \right)
    \right].
\end{eqnarray}
The redistribution of occupation will take place only from the higher-energy 
towards the lower-energy molecular state.

\section{Steady-state equations}
\label{app:dmf}

The equations for the elements $n_i$ corresponding to each of the eight two-hole
quantum molecular states are:
\begin{eqnarray}
    0 & = & 
    \left( \Gamma_i^{(in)}(\Downarrow,\Downarrow) 
    + \Gamma_i^{(in)}(\Uparrow,\Downarrow)\right) n_{L\Downarrow}
    + \left( \Gamma_i^{(in)}(\Downarrow,\Uparrow) 
    + \Gamma_i^{(in)}(\Uparrow,\Uparrow)\right) n_{L\Uparrow} \nonumber\\
    &-& \sum_{\sigma_1}\sum_{\sigma_2} \Gamma_i^{(out)}(\sigma_1,\sigma_2) n_i
    + \sum_{k = i+1}^{8} \Gamma^{(SF)}_{ki} n_k -
    \left(\sum_{j=1}^{i-1}\Gamma^{(SF)}_{ij}\right)n_i,
\end{eqnarray}
with the first two terms describing the hole tunneling from the source,
the third term treating the hole tunneling into the drain, and the last two
accounting for the spin-flip relaxation.
For the single-hole states we have:
\begin{eqnarray}
0 &=& - \sum_{i=1}^8 \left( \Gamma_i^{(in)}(\Downarrow,\Downarrow) 
    + \Gamma_i^{(in)}(\Uparrow,\Downarrow)\right) n_{L\Downarrow}
    + \sum_{i=1}^8 \left( \Gamma_i^{(out)}(\Downarrow,\Downarrow) 
    + \Gamma_i^{(out)}(\Uparrow,\Downarrow)\right) n_{i} + \frac{n_{L\Uparrow}}{T_1} ,\\
0 &=& - \sum_{i=1}^8 \left( \Gamma_i^{(in)}(\Downarrow,\Uparrow) 
    + \Gamma_i^{(in)}(\Uparrow,\Uparrow)\right) n_{L\Uparrow}
    + \sum_{i=1}^8 \left( \Gamma_i^{(out)}(\Downarrow,\Uparrow) 
    + \Gamma_i^{(out)}(\Uparrow,\Uparrow)\right) n_{i} - \frac{n_{L\Uparrow}}{T_1}.
\end{eqnarray}
This system of equations is solved algebraically with a subsidiary condition
\begin{equation}
    \sum_{i=1}^{8} n_i  + n_{L\Uparrow} + n_{L\Downarrow} = 1.
\end{equation}


\begin{thebibliography}{99}
\bibitem{Scappucci2020}
G. Scappucci, C. Kloeffel, F. A. Zwanenburg, D. Loss,
M. Myronov, J.-J. Zhang, S. De Franceschi, G. Katsaros,
and M. Veldhorst, arxiv:2004.08133v1 (2020).

\bibitem{Burkard2008}
G. Burkard, Nature Materials {\bf 7}, 100 (2008).

\bibitem{Fischer2008}
J. Fischer, W. A. Coish, D. V. Bulaev, and D. Loss,
Phys. Rev. B {\bf 78}, 155329 (2008).

\bibitem{Bulaev2007}
D. V. Bulaev and D. Loss,
Phys. Rev. Lett. {\bf 98}, 097202 (2007).

\bibitem{Venitucci2019}
B. Venitucci and Y.-M. Niquet,
Phys. Rev. B {\bf 99}, 115317 (2019).

\bibitem{Voisin2016}
B. Voisin, R. Maurand, S. Barraud, M. Vinet,
X. Jehl, M. Sanquer, J. Renard, and S. De Franceschi,
Nano Lett. {\bf 16}, 88 (2016).

\bibitem{Maurand2016}
R. Maurand, X. Jehl, D. Kotekar-Patil, A. Corna,
H. Bohuslavskyi, R. Lavi{\'{e}}ville, L. Hutin, S. Barraud, 
M. Vinet, M. Sanquer, and S. De Franceschi,
Nature Communications {\bf 7}, 13575 (2016).

\bibitem{Watzinger2018} 
H. Watzinger, J. Kuku{\v{c}}ka, L. Vuku{\v{s}}i{\'{c}}, 
F. Gao, T. Wang, F. Sch{\"{a}}ffler, J.-J. Zhang, and G. Katsaros,
Nature Communications {\bf 9}, 3902 (2018).

\bibitem{Bogan2019-2}
S. Studenikin, M. Korkusinski, M. Takahashi, J. Ducatel, A. Padawer-Blatt, 
A. Bogan, D. G. Austing, L. Gaudreau, P. Zawadzki, A. Sachrajda, Y. Hirayama, 
L. Tracy, J. Reno, and T. Hargett, 
Nature Communications Physics {\bf 2}, 159 (2019).

%Si planar
\bibitem{Li2015}
R. Li, F. E. Hudson, A. S. Dzurak, and A. R. Hamilton,
Nano Lett. {\bf 15}, 7314 (2015).

%Ge/Si huts

\bibitem{Gao2020}
F. Gao,  J.‐H. Wang,  H. Watzinger,  H. Hu,  M. J. Rancic,
J.‐Y. Zhang,  T. Wang,  Y. Yao,  G.‐L. Wang,  J. Kuku{\v{c}}ka, 
L. Vuku{\v{s}}i{\'{c}},  C. Kloeffel,  D. Loss,  F. Liu,  G. Katsaros,
and J.‐J. Zhang,
Adv. Mater. {\bf  32}, 1906523 (2020).

\bibitem{Watzinger2016}
H. Watzinger, C. Kloeffel, L. Vuku{\v{s}}i{\'{c}}, 
M. D. Rossell. V. Sessi, J. Kuku{\v{c}}ka, R. Kirchschlager,
E. Lausecker, A. Truhlar, M. Glaser, A. Rastelli, A. Fuhrer,
D. Loss, and G. Katsaros,
Nano Letters {\bf 16}, 6879 (2016).

\bibitem{Li2018}
Y. Li, S.-X. Li, F. Gao, H.-O. Li, G. Xu, K. Wang, D. Liu, G. Cao,
M. Xiao, T. Wang, J.-J. Zhang, G.-C. Guo, and G.-P. Guo,
Nano Letters {\bf 18}, 2091 (2018).

%Ge-Si core shell
\bibitem{Hu2007}
Y. Hu, H. O. H. Churchill, D. J. Reilly, J. Xiang, C. M. Lieber,
and C. M. Marcus,
Nature Nanotechnology {\bf 2}, 622 (2007).

\bibitem{Hu2012}
Y. Hu, F. Kuemmeth, C. M. Lieber, and C. M. Marcus,
Nature Nanotechnology {\bf 7}, 47 (2012).

\bibitem{Ares2013}
N. Ares, V. N. Golovach, G. Katsaros, M. Stoffel, F. Fournel, 
L. I. Glazman, O. G. Schmidt, and S. De Franceschi,
Phys. Rev. Lett. {\bf 110}, 046602 (2013).

%planar germanium
\bibitem{Hendrickx2020}
N. Hendrickx, D. Franke, A. Sammak, G. Scappucci, and
M. Veldhorst, 
Nature {\bf 577}, 487 (2020).

\bibitem{Hendrickx2019}
N.W. Hendrickx, W.I.L. Lawrie, L. Petit, A. Sammak, G. Scappucci, 
and M. Veldhorst, arXiv:1912.10426.

%GaAs/AlGaAs
\bibitem{Wang2016}
D. Q. Wang, O. Klochan, J.-T. Hung, D. Culcer, I. Farrer, D. A.
Ritchie, and A. R. Hamilton,
Nano Lett. {\bf 16}, 7685 (2016).

\bibitem{Wang2016-2}
D. Q. Wang, A. R. Hamilton, I. Farrer, D. A. Ritchie,
and O. Klochan,
Nanotechnology {\bf 27}, 334001 (2016).
%--
\bibitem{Bogan2019}
A. Bogan, S. Studenikin, M. Korkusinski, L. Gaudreau, P. Zawadzki, A. Sachrajda, 
L. Tracy, J. Reno, and T. Hargett,
Nature Communications Physics {\bf 2}, 17 (2019).

\bibitem{Tracy2014}
L. A. Tracy, T. W. Hargett, and J. L. Reno,
Appl. Phys. Lett. {\bf 104}, 123101 (2014).

\bibitem{Bogan2017}
A. Bogan, S. A. Studenikin, M. Korkusinski, G. C. Aers, L. Gaudreau, P. Zawadzki, 
A. S. Sachrajda, L. A. Tracy, J. L. Reno, and T. W. Hargett,
Phys. Rev. Lett. {\bf 118}, 167701 (2017).

\bibitem{Bogan2018}
A. Bogan, S. Studenikin, M. Korkusinski, L. Gaudreau, P. Zawadzki, A. S. Sachrajda, 
L. Tracy, J. Reno, and T. Hargett.
Phys. Rev. Lett. {\bf 120}, 207701 (2018).

\bibitem{Bulaev2005}
D. V. Bulaev and D. Loss, Phys. Rev. Lett. {\bf 95}, 076805 (2005).

\bibitem{NadjPerge2010}
S. Nadj-Perge, S. M. Frolov, J. W. W. van Tilburg, J. Danon, 
Yu. V. Nazarov, R. Algra, E. P. A. M. Bakkers, and L. P. Kouwenhoven,
Phys. Rev. B {\bf 81}, 201305(R) (2010).

\bibitem{Nichol2015}
J. M. Nichol, S. P. Harvey, M. D. Shulman, A. Pal, V. Umansky, 
E. I. Rashba, B. I. Halperin, and  A. Yacoby 
Nature Communications {\bf 6}, 7682 (2015).


\bibitem{Ciorga2000}
M. Ciorga, A. S. Sachrajda, P. Hawrylak, C. Gould, P. Zawadzki, 
S. Jullian, Y. Feng, and Z. Wasilewski,
Phys. Rev. B {\bf 61}, 16315(R) (2000).

\bibitem{Ono2002}
K. Ono, D. G. Austing, Y. Tokura, and S. Tarucha,
Science  {\bf 297}, 1313 (2002).

\bibitem{Taylor2007}
J. M. Taylor, J. R. Petta, A. C. Johnson, A. Yacoby, C. M. Marcus, 
and M. D. Lukin, 
Phys. Rev. B {\bf 76}, 035315 (2007), 

\bibitem{Petta2005}
J. R. Petta, A. C. Johnson, J. M. Taylor, E. A. Laird, A. Yacoby, 
M. D. Lukin, C. M. Marcus, M. P. Hanson, and A. C. Gossard,
Science {\bf 309}, 2180 (2005).

\bibitem{Petta2010}
J. R. Petta, H. Lu, and A. C. Gossard, 
Science {\bf 327}, 669 (2010).

\bibitem{Studenikin2012}
S. A. Studenikin, G. C. Aers, G. Granger, L. Gaudreau, A. Kam, 
P. Zawadzki, Z. R. Wasilewski, and A. S. Sachrajda, 
Phys. Rev. Lett. {\bf 108}, 226802 (2012).

\bibitem{Korkusinski2017}
M. Korkusinski, S. A. Studenikin, G. Aers, G. Granger, A. Kam, 
and A. S. Sachrajda, 
Phys. Rev. Lett. {\bf 118}, 067701 (2017).

\bibitem{Kyriakidis2002}
J. Kyriakidis, M. Pioro-Ladriere, M. Ciorga, A. S. Sachrajda, and P. Hawrylak,
Phys. Rev. B {\bf 66}, 035320 (2002).

\bibitem{Szumniak2012}
P. Szumniak, S. Bednarek, B. Partoens, and F. Peeters,
Phys. Rev. Lett. {\bf 109}, 107201 (2012).

\bibitem{Szumniak2013}
P. Szumniak, S. Bednarek, J. Pawlowski, and B. Partoens,
Phys. Rev. B {\bf 87}, 195307 (2013).

\bibitem{Luttinger1955}
J. M. Luttinger and W. Kohn, Phys. Rev. {\bf 97}, 869 (1955).

\bibitem{Luttinger1956}
J. M. Luttinger, Phys. Rev. {\bf 102}, 1030 (1956).

\bibitem{Dresselhaus1955}
G. Dresselhaus, Phys. Rev. {\bf 100}, 580 (1955).

\bibitem{Cardona1988}
M. Cardona, N. E. Christiensen, and G. Fasol, Phys. Rev. B {\bf 38}, 1806 (1988).

\bibitem{Burkard99}
G. Burkard, D. Loss, and D. P. DiVincenzo, Phys. Rev. B {\bf 59}, 2070 (1999).

\bibitem{Calderon06}
M. Calderon, B. Koiller, and S. D. Sarma, Phys. Rev. B {\bf 74}, 045310 (2006).

\bibitem{Hu00}
X. Hu and S. D. Sarma, Phys. Rev. A {\bf 61}, 062301 (2000).

\bibitem{Wiel06}
W. van der Wiel, M. Stopa, T. Kodera, T. Hatano, and S. Tarucha, 
New J. Phys. {\bf 8}, 28 (2006).

\bibitem{Hatano08}
T. Hatano, S. Amaha, T. Kubo, Y. Tokura, Y. Nishi, Y. Hirayama, and S. Tarucha, 
Phys. Rev. B {\bf 77}, 241301 (2008).

\bibitem{Gimenez2007}
I. Puerto Gimenez, M. Korkusinski, and P. Hawrylak,
Phys. Rev. B {\bf 76}, 075336 (2007).

\bibitem{Buonacorsi20}
B. Buonacorsi, M. Korkusinski, B. Khromets, and J. Baugh,
arXiv:2012.10512 (2020).


\bibitem{Pfund2007}
A. Pfund, I. Shorubalko, K. Ensslin, and R. Leturcq,
Phys. Rev. B {\bf 76}, 161308(R) (2007).

\bibitem{Ciorga2002}
M. Ciorga, A. Wensauer, M. Pioro-Ladriere, M. Korkusinski, J. Kyriakidis, A. S. Sachrajda, 
and P. Hawrylak
Phys. Rev. Lett. {\bf 88}, 256804 (2002).

\bibitem{Qassemi2009}
F. Qassemi, W. A. Coish, and F. K. Wilhelm,
Phys. Rev. Lett. {\bf 102}, 176806 (2009).


\end{thebibliography}
\end{document}